\documentclass[nofootinbib,prd,twocolumn,showpacs,showkeys,preprintnumbers]{revtex4-1}
\usepackage{hyperref,amssymb,amsmath,mathrsfs,bm,graphicx}
\usepackage{multirow}
\usepackage{adjustbox}
\usepackage{array}
\usepackage{dblfloatfix}
\usepackage{placeins}
\usepackage{float}
\usepackage[dvipsnames]{xcolor}
\begin{document}

\title {Quasi normal modes of a Casimir--like traversable wormhole through the semi-analytical WKB approach.}

\author{R. Avalos}
\email{ravalos@usfq.edu.ec}
\affiliation{Departamento de F\'isica, Colegio de Ciencias e Ingenier\'ia, Universidad San Francisco de Quito,  Quito 170901, Ecuador.\\}

\author{E. Contreras }
\email{econtreras@usfq.edu.ec}
\affiliation{Departamento de F\'isica, Colegio de Ciencias e Ingenier\'ia, Universidad San Francisco de Quito,  Quito 170901, Ecuador.\\}

\begin{abstract}
In this work, we implement the semi-analytical WKB method to explore the behaviour of a scalar field on a traversable wormhole space--time with a Casimir--like complexity reported in Eur. Phys. J. C  82, 420 (2022). We estimate the error in the computation of the quasi--normal frequencies of the scalar field at each order in the WKB and show that the order with the best accuracy is not unique. We compute the value of the quasinormal frequencies for the smallest estimated error. As an aside,  we find that the imaginary part of the quasinormal modes frequencies approach to zero for the fundamental mode mimicking the so--called quasi--resonance observed for massive scalar fields in the Reissner--Nordstr\"{o}m background.
\end{abstract}

\keywords{QNM}

\maketitle
\section{Introduction}
Undoubtedly, the study of traversable wormholes (TW's) \cite{Morris:1988cz,Morris:1988tu,Alcubierre:2017pqm,Visser:1995cc,Lobo:2005us,Garattini:2019ivd,Stuchlik:2021guq,Bronnikov:2021ods,Blazquez-Salcedo:2021udn,Churilova:2021tgn,Konoplya:2021hsm,Tello-Ortiz:2021kxg,Bambi:2021qfo,Capozziello:2020zbx,Blazquez-Salcedo:2020czn,Berry:2020tky,Maldacena:2020sxe,Garattini:2020kqb}  is an attractive research area not only for the intriguing features they encode regarding the possibility of interstellar travels but for their potential as black hole mimickers as has been established in recent years \cite{Bueno:2017hyj,Cardoso:2016oxy,Konoplya:2016hmd}. On the one hand, the analysis of how a TW responds to perturbations allows one to extract useful information about its similitude with the ringdown phase of a black hole through its quasinormal modes (QNM) spectrum. On the other hand, the behaviour of QNM frequencies can be related to the stability of the perturbing field itself as its imaginary part, $Im(\omega)$, is related to the damping factor associated with the loss of energy through gravitational radiation. Indeed, depending on the sign of
$Im(\omega)$ (see Eq. (5)) the perturbation either grows exponentially leading to instabilities or decreases monotonously. 

The study of the QNM can be carried out by different methods (for an incomplete list see \cite{Konoplya:2022tvv,Churilova:2021nnc,Konoplya:2020jgt,Konoplya:2020bxa,Konoplya:2019nzp,Rincon:2021gwd,Panotopoulos:2020mii,Rincon:2020cos,Rincon:2020iwy,Rincon:2020pne,Xiong:2021cth,Zhang:2021bdr,Panotopoulos:2019gtn,Lee:2020iau,Churilova:2019qph,Oliveira:2018oha,Blazquez-Salcedo:2018ipc,Panotopoulos:2017hns} and references therein, for example). Nevertheless, in this work, we shall implement the semi--analytical WKB approximation reported in \cite{Konoplya:2019hlu}. In particular, we analyse the QNM for the TW reported in \cite{Avalos:2022tqg} that fulfills all the basic requirements of a TW, namely, flaring out condition, small tidal forces, reasonable time to traverse the throat and, an arbitrarily small amount of exotic matter. It is worth mentioning that, although the solution in \cite{Avalos:2022tqg} depends on a free parameter which is bounded in accordance to the above mentioned requirements, it is our main goal here to analyse the possibility of refining such an interval by exploring the stability of the scalar perturbation. It should be emphasized that this is the first study on QNM for this solution in the literature.

This work is organized as follows. In the next section we introduce the computation of the QNM frequencies by the WKB approximation. Next, in section \ref{model}, we review the main aspects of the TW with Casimir--like complexity reported in \cite{Avalos:2022tqg}. The results obtained in this work are entirely
contained in section \ref{results} where we discuss their impact and scope. It is worth mentioning that all of our computations were performed with the Mathematica notebook publicly available in https://goo.gl/nykYGL. In the last section we conclude the work. 

\section{QNM by the WKB approximation}\label{QNMi}
Let us start by considering the standard line element of a TW in spherical coordinates given by \cite{Morris:1988cz}
\begin{equation}\label{metric}
ds^{2}=-e^{2\phi} dt^2 +d r^{2}/(1-b/r)+r^{2}(d\theta^{2}+\sin^{2}\theta d\phi^{2}),
\end{equation}
where $\phi=\phi(r)$ is the redshift function which encodes information on the radial tidal force and
$b=b(r)$ is the shape function which enciphers the main features of the throat of the TW.

It is well known that we can perturb a TW by considering small deviations in the space--time background or by studying the evolution of test fields in the underlying geometry. Remarkably, whatever the path we follow, the behaviour of the perturbation field is governed by a 
Schr\"{o}dinger--like equation, namely
\begin{eqnarray} \label{schrodinger}
\bigg( \frac{d^2}{dr_*^2} +\omega ^2 -V(r_*) \bigg)\chi(r_*)=0,
\end{eqnarray}
where
\begin{eqnarray}
r_*(r)=\int_{r_0}^r \frac{1}{\sqrt{1-b(r')/r'}}dr',
\end{eqnarray}
is the tortoise coordinate
and $V(r)$ is the effective potential whose functional profile depends on the particular metric given by Eq. (\ref{metric}). It is worth noticing that, as $r_*\in(-\infty, \infty)$,  
the throat of the wormhole is at $r_*=0$ and
the asymptotically flat region far from the throat corresponds to $r_*=\pm\infty$. In this manuscript, we shall perturb the TW geometry with a massless scalar field, so the potential corresponds to 
\begin{eqnarray} \label{veff}
V_L (r) = e^{2\phi} \bigg(\frac{L(L+1)}{r^2}-\frac{r b' -b}{2r^3} +\frac{\phi'}{r} \bigg(1-\frac{b}{r} \bigg) \bigg),
\end{eqnarray}
where $L$ represents the multipole number. 

We are looking for solutions of
(\ref{schrodinger}) such that the wave is purely out--going when we approach to infinity, namely
\begin{eqnarray} \label{bc}
\chi (r_*) \sim C_{\pm}\exp (\pm i \omega r_*), \,\,\,\,\, r_*\rightarrow \pm \infty.
\end{eqnarray}
In this case, the solution is called a QNM with $\omega=Re(\omega)+iIm(\omega)$. It is clear that the QNM frequencies encode valuable information about the evolution of the perturbation field. Indeed, it is well--known that, on the one hand, the real part of $\omega$ gives information about the oscillation of the signal and, on the other hand, the imaginary part $Im(\omega)$ accounts for the damping factor associated to the energy loss by  gravitational radiation. Even more,
when $Im(\omega)>0$, the perturbation is unstable given that
the growth of the
perturbation field is exponential so to ensure the stability we demand that $Im(\omega)<0$. 

Although the QNM frequencies can be computed by numerical methods, in this work, we shall take advantage of the fact that Eq. (\ref{schrodinger}) coincides formally with
the one-dimensional Schr\"{o}dinger equation which is the preferred arena to implement the well--known WKB approximation. It is worth mentioning that this strategy was introduced 
in Ref. \cite{Schutz:1985km} and has been refined to higher orders in \cite{Konoplya:2019hlu} to deal with perturbations of black hole geometries. As such, the same strategy has been adopted to work the problem of TW with bell--shaped potentials in
\cite{Churilova:2019qph}. In any case, the formula for the computation of the QNM frequencies to the $k^{th}$ order in perturbation reads
\begin{eqnarray} \label{WKB}
i \frac{\omega ^2 -V_0}{\sqrt{-2V_0''}}-\sum_{j=2}^{k}\Lambda_j=k+\frac{1}{2},
\end{eqnarray}
where $V_0$ and $V_0''$ stand for the maximum of the potential and its second derivative respect to the tortoise coordinate, respectively. In the case of TW with bell--shaped potentials (see Fig. \ref{potential}), the maximum is located at the throat ($r=r_0$, $r_*=0$). Finally, $\Lambda_j$ contains all the higher order corrections and can be found in \cite{Konoplya:2019hlu}. In Eq. (\ref{WKB}), $k$ is the order of the WKB. It should be emphasized that, an increasing in $k$ does not necessarily lead to a better approximation of the quasinormal frequencies . To be more precise, the order $k$ leading to the best value of $\omega$ is not unique but could depend on the pair $(n,L)$ chosen. For that reason, in this work we shall implement the following strategy for the computation of the best $\omega$ for each order:\\ 
1. We estimate the accuracy by using \cite{Konoplya:2019hlu}
\begin{equation} \label{Deltak}
\Delta_k= \frac{| \omega_{k+1}-\omega_{k-1} |}{2},
\end{equation}
which, as discussed in \cite{Konoplya:2019hlu}, is usually greater than the error, namely
\begin{eqnarray}\label{errorD}
\Delta_{k}\geq |\omega-\omega_{k}|.
\end{eqnarray}
with $\omega$ the accurate value of the quasinormal frequency. From (\ref{errorD}) we see that as $\Delta_{k}\to0$ for some $k$, the accuracy of the WKB increases. \\
2. We compute $Im(\omega)$ and $Re(\omega)$ with the smallest $\Delta_{k}$ associated to a given pair $(n,L)$.\\
\\
In the next section we briefly introduce the TW solution in which we will focus our analysis

\section{The model}\label{model}
The model we are interesting in here corresponds to the TW with Casimir--like complexity supported by a arbitrarily small amount of exotic matter which metric is given by
\cite{Avalos:2022tqg}
\begin{eqnarray}
\phi(r)&=& \ln \bigg( \frac{c_0 r}{c_0 r +r_0} \bigg) \label{phiCasimir}\\
b(r)&=&\frac{2r_0}{c_{0}} +\bigg( 1-\frac{2}{c_{0}} \nonumber\\ 
&&-\frac{4 \big[4-2 c_{0}+
\ln\left(1+\frac{1}{c_{0}}\right)(c_{0}^{2}-2c_{0}-3)
 \big]}{ c_{0} \big[-3-4c_{0} +4c_0(c_0+1) \ln\big(1+\frac{1}{c_{0}}\big)\big]} \bigg) \frac{r_0^2}{r} \nonumber\\ 
&&+\frac{4 \big[4-2 c_{0}+
\ln\left(1+\frac{1}{c_{0}}\right)(c_{0}^{2}-2c_{0}-3)
 \big]}{ c_{0} \big[-3-4c_{0} +4c_0(c_0+1) \ln\big(1+\frac{1}{c_{0}}\big)\big]}\frac{r_0^3}{r^2},\label{bCasimir}
\end{eqnarray}
where $c_0$ is the only free parameter which is restricted to the values 
\begin{equation} \label{brange}
   0.950679<c_{0}<4.86215.
\end{equation}
It is worth mentioning that, the interval above ensures (see \cite{Avalos:2022tqg} for details): i) a finite time of travel while traversing the wormhole, ii) a reasonable values for the radial tidal force and iii) an arbitrarily small amount of exotic matter. Although the solution has been analysed in some detail in \cite{Avalos:2022tqg}, in the next section we complement it by exploring its response to scalar perturbations.

\section{Results and Discussion}\label{results}
In this section, we show and discuss the response of the Casimir--like TW in \cite{Avalos:2022tqg}
to scalar perturbations.

\begin{figure*}[htb!]
\centering
\includegraphics[width=0.4\textwidth]{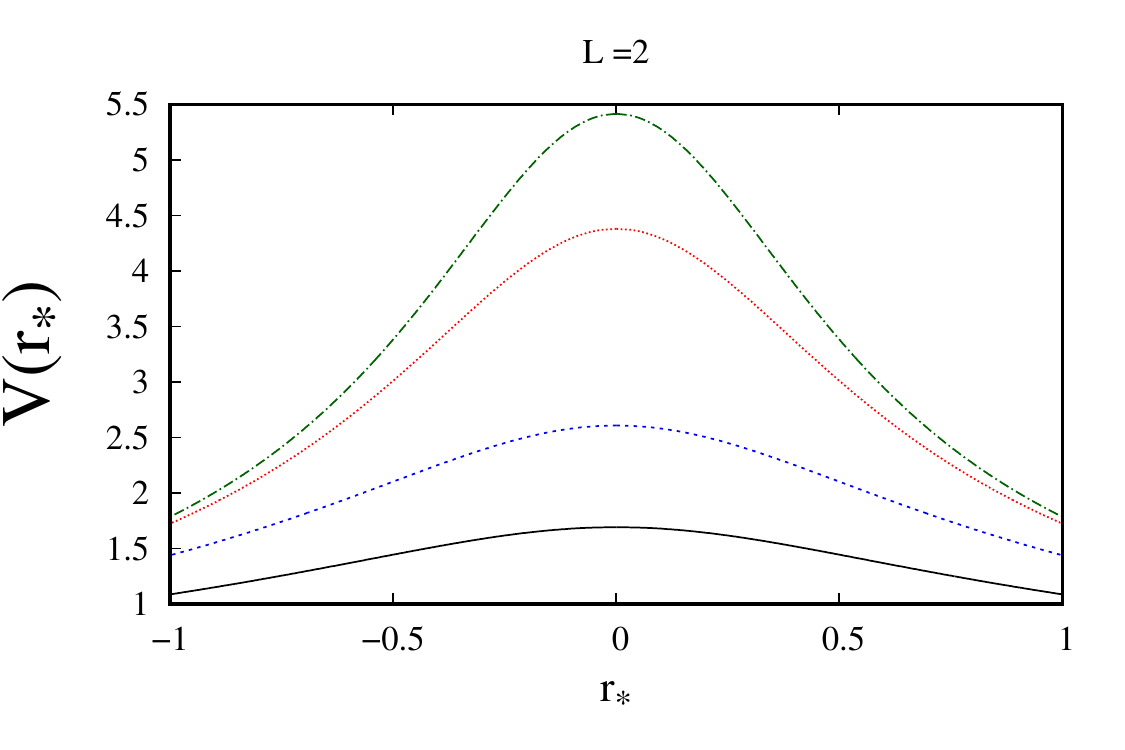}  \
\includegraphics[width=0.4\textwidth]{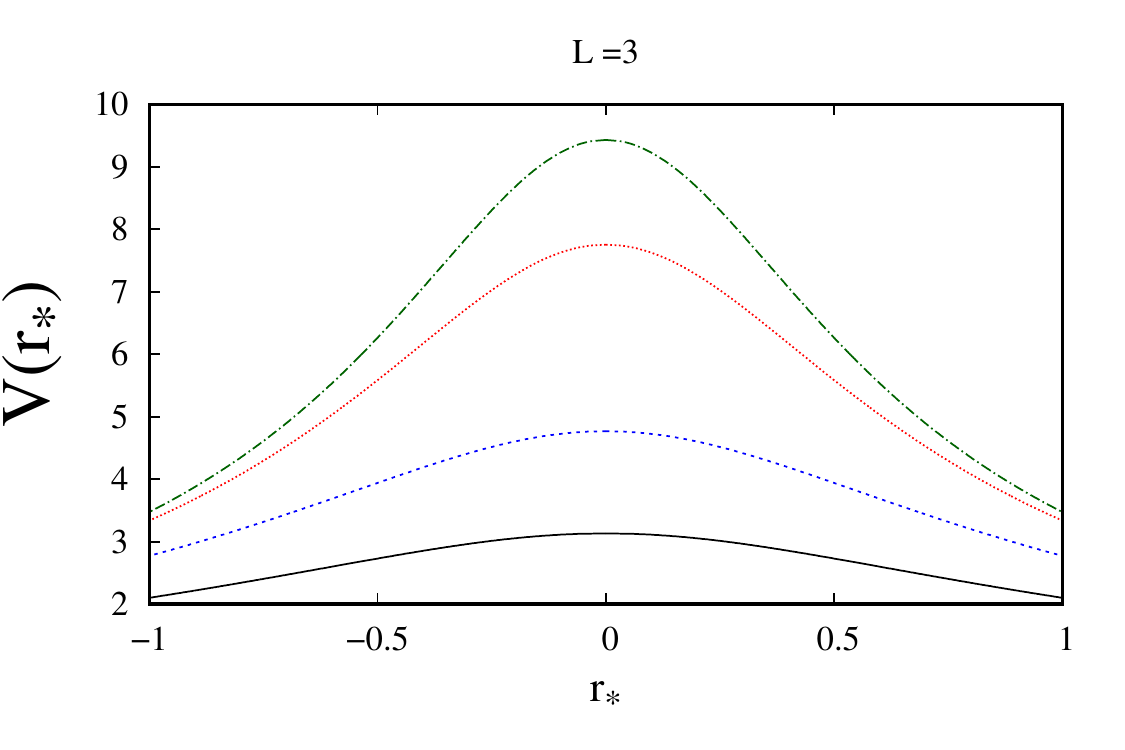}  \
\includegraphics[width=0.4\textwidth]{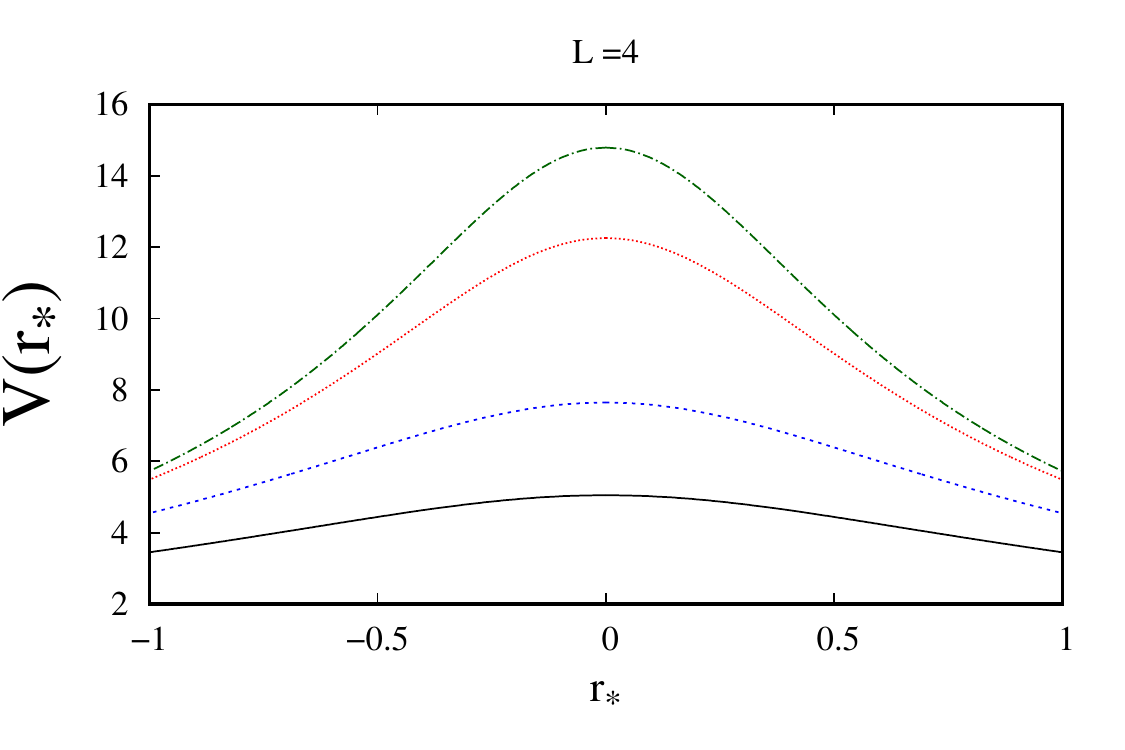}  \
\includegraphics[width=0.4\textwidth]{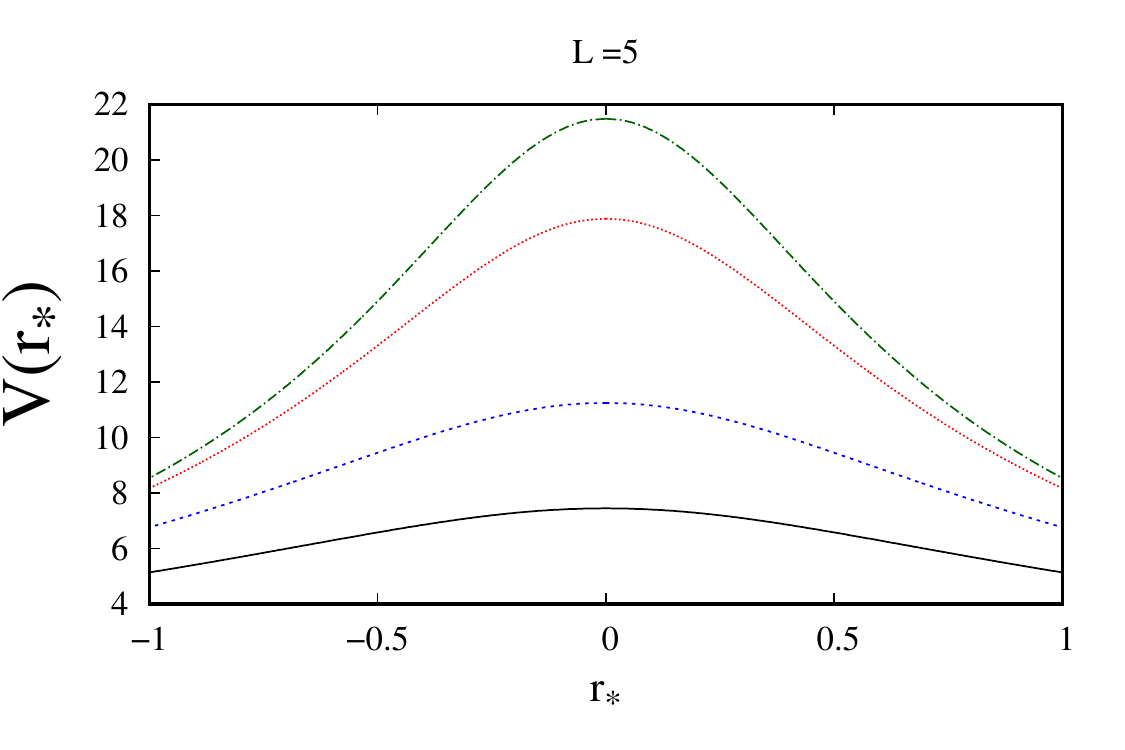}  \
\includegraphics[width=0.4\textwidth]{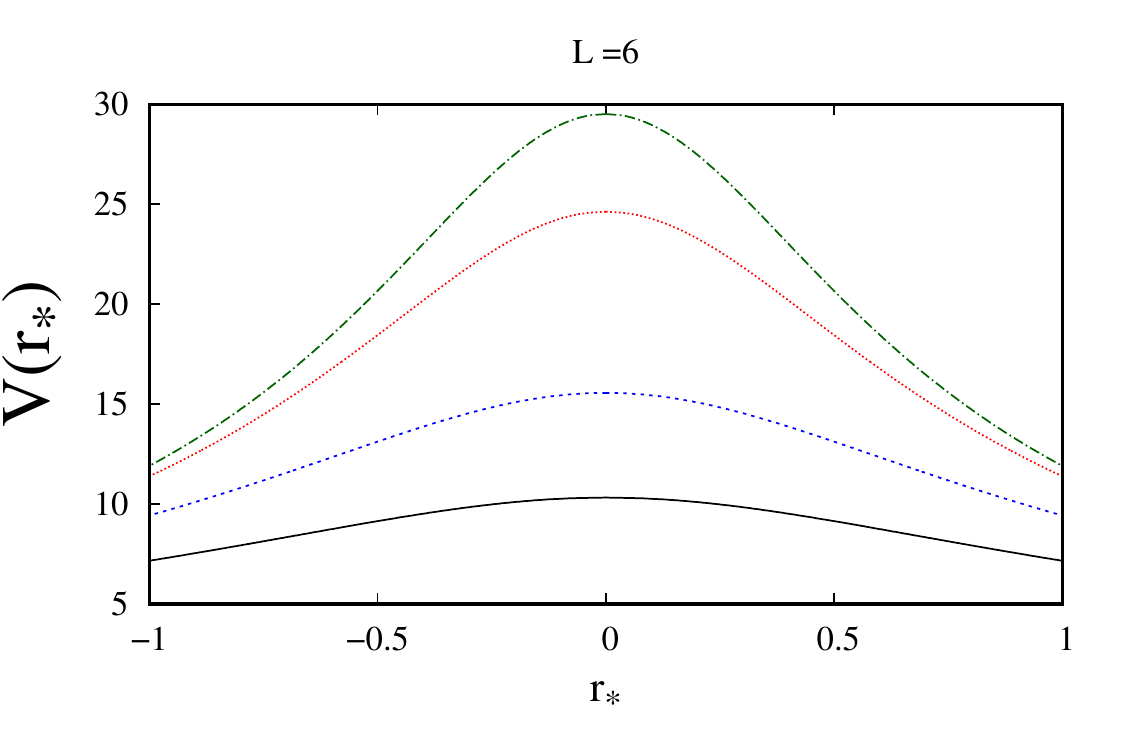}  \
\includegraphics[width=0.4\textwidth]{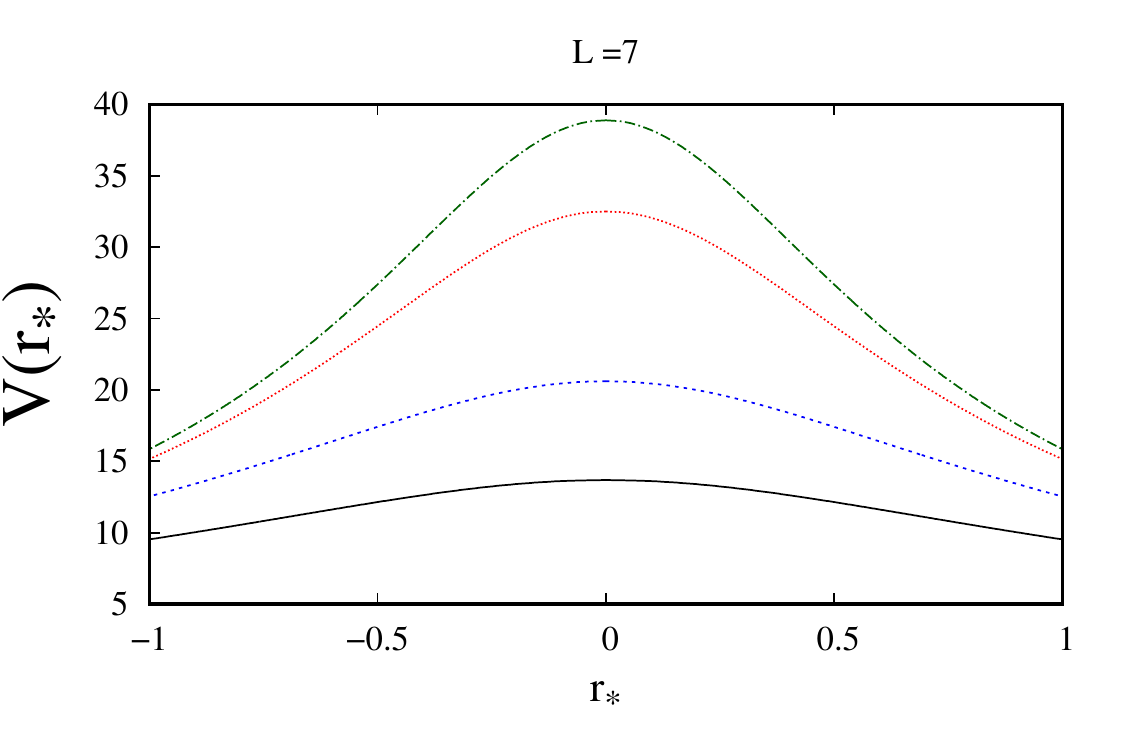}  \
\includegraphics[width=0.4\textwidth]{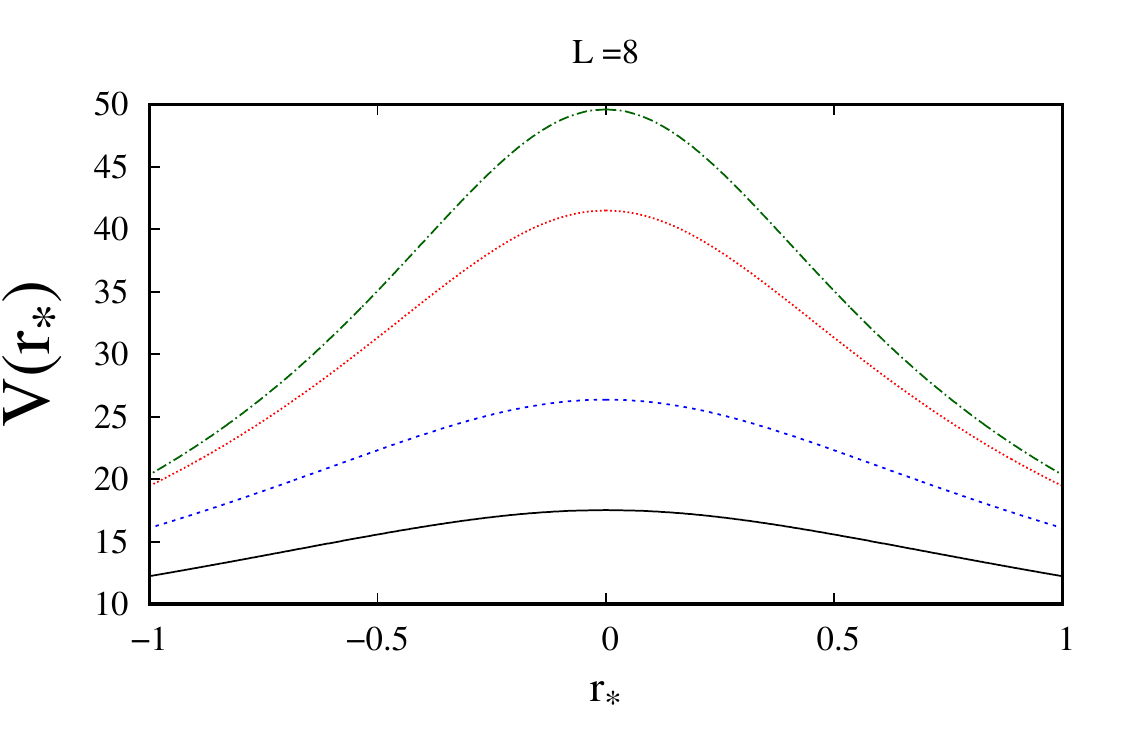}  \
\includegraphics[width=0.4\textwidth]{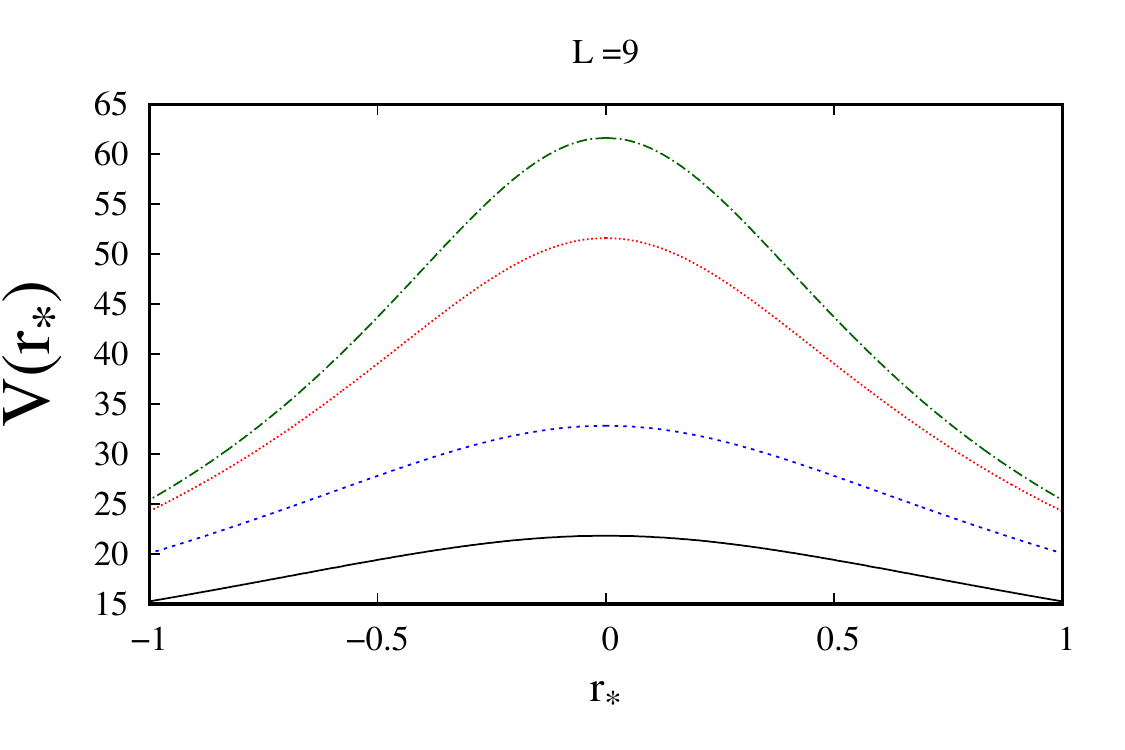} \
\includegraphics[width=0.4\textwidth]{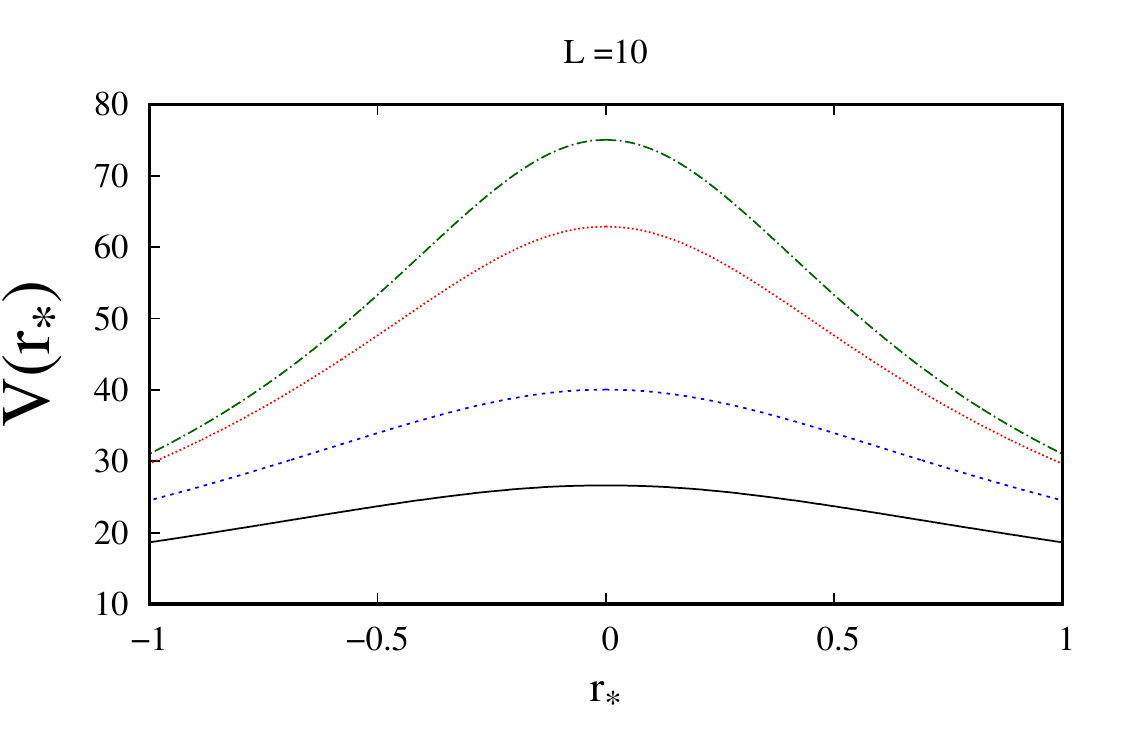}  \
\caption{\label{potential}
Effective potential of the QNM for the wormhole as a function of the tortoise coordinate $r_*$ for different values of $L$ and $c_0=0.96$ (black line), $c_0=1.5$ (blue line), $c_0=3$ (red line), $c_0=4.5$ (green line). }
\end{figure*}

In Fig. (\ref{potential}) it is shown the effective potential given by Eq. \eqref{veff} parameterized by $c_{0}$ as a function of the tortoise coordinate $r_*$ for different values of $L$. Note that the potential reaches a maximum at the throat ($r_*=0$) as required for the implementation the WKB method \footnote{In the case that there were other local maximum points this numerical method will not work.}. The behaviour of the potential with respect to $L$ is clear: as $L$ increases the height of the potential also increases. Notice that the maximum values of the potential for $L=10$ are around one order of magnitude larger than the values for $L=2$. Similarly, the barrier also increases as the free parameter $c_0$ grows.\\

In Tables \ref{table:1} and \ref{table:2} we show the values of $\Delta_k$ and $Im(\omega_{k})$ for $k\in[2,7]$, $n=0,1,2$ and $L=0,1,2,...,10$ for different values of the free parameter $c_{0}$. Note that, the smallest $\Delta_{k}$ for each pair $(n,L)$ has been highlighted in red color. Note that as $c_0$ grows, the smallest $\Delta_{k}$ appears when $k$ increases. Similarly, for a fixed $c_0$, $\Delta_k$ reach its minimum value for bigger $k$ as $L$ increases. However, it should be emphasized that for small $L$  the minimum value reached by $\Delta_{k}$ is considerably greater than its value for high $L$. For example, for $c_{0}=0.96$ and $n=0$, the smallest error is obtained for $k=4$ but $\Delta_{4}=0.6518$ for $L=0$ and $\Delta_{4}=0.0068$ for $L=10$, which corresponds to a ratio of two orders of magnitude. In this regard, we should consider the quasinormal frequency obtained for $L=0$ as not accurate. A possible consequence of such a lack of accuracy, its that the computation of $Im(\omega)$ could lead to a positive number (indicated with a dash in Table \ref{table:2}) which does not necessarily indicate
an instability of the scalar field. What is more, the WKB cannot be used for unstable modes in principle so that
positive values are meaningless. Indeed, given that the potential is positive everywhere (as shown in Fig. (\ref{potential})) the stability of the scalar field is ensured.  It worth mentioned that we have computed $\Delta_{k}$ for $k=8,9,10,11,12$ but the error increases considerably. For this reason we have not shown this data here.

\begin{table*}[t]
\centering
\begin{adjustbox}{width=11cm,center}
\begin{tabular}{|m{4em}|m{2em}|m{2em}|m{4em}|m{4em}|m{4em}|m{4em}|m{4em}|m{4em}|} 
 \hline
 $c_0$ & $L$ & $n$ &$\Delta_{2}$ & $\Delta_{3}$ &$\Delta_{4}$ &$\Delta_{5}$&$\Delta_{6}$&$\Delta_{7}$ \\  
\hline
0.96& 0 & 0 & 0.8550& 0.9151 & \textcolor{red}{0.6518} & 0.9289 & 1.0371 & 2.7327\\ 
& & 1 & \textcolor{red}{1.8959}& 2.4788 & 2.5087 & 1.9295 & 4.5556 & 4.9325 \\ 
& & 2 & \textcolor{red}{2.9752}& 5.1475& 6.9424 & 10.5522 & 23.9017 & 43.2611 \\
& 1 & 0 & \textcolor{red}{0.06815}& 0.7068 & 0.9036 & 0.6302 & 0.7020 & 2.0710 \\ 
& & 1 & 1.7753& 2.0336 & \textcolor{red}{0.7251} & 0.8488 & 2.0417 & 3.9368  \\ 
& & 2 & 2.9641& 4.4068& \textcolor{red}{1.2328} & 6.1848 & 5.7790 & 19.4865  \\
& 2 & 0 & \textcolor{red}{0.3347}& 0.4074 & 0.6926 & 1.2441 & 1.9601 & 2.5636 \\ 
& & 1 & 1.5434& \textcolor{red}{1.5309} & 2.0627 & 2.3138 & 2.8719 & 5.8893  \\ 
& & 2 & 2.6514& 3.0877& 1.3139 & \textcolor{red}{0.9384} & 2.7061 & 4.9520  \\
& 3 & 0 & \textcolor{red}{0.1713}& 0.2018 & 0.3511 & 0.9177 & 1.7722 & 3.4042\\ 
& & 1 & 1.4050& \textcolor{red}{1.0113} & 2.0561 & 2.5588 & 4.4838 & 7.7817  \\ 
& & 2 & \textcolor{red}{2.3602}& 2.4098& 3.0993 & 4.0485 & 6.1083 & 10.7085 \\
& 4 & 0 & 0.1094 & \textcolor{red}{0.1048} & 0.1642 & 0.4061 & 1.0116 & 2.0633\\
& & 1 & 0.6171 & \textcolor{red}{0.5983} & 0.9170 & 1.8061 & 3.6314 & 6.6879 \\ 
& & 2 & 2.2306 & \textcolor{red}{1.7326} & 3.0042 & 3.7870 & 6.6206 & 12.4492 \\
& 5 & 0 & 0.0791 & \textcolor{red}{0.0602} & 0.0799 & 0.1618 & 0.4127 & 0.9598 \\ 
& & 1 & 0.4165 & \textcolor{red}{0.3579} & 0.4890 & 1.0558 & 2.4535 & 4.3869 \\ 
& & 2 & 1.3130 & \textcolor{red}{1.1551} & 1.5898 & 2.6949 & 5.2246 & 9.5597 \\
& 6 & 0 & 0.0610& \textcolor{red}{0.0380} & 0.0423 & 0.0725 & 0.1629 & 0.4025 \\ 
& & 1 & 0.3177& \textcolor{red}{0.2284} & 0.2707 & 0.5284 & 1.4187 & 2.4720  \\ 
& & 2 & 0.9260& \textcolor{red}{0.7599}& 0.9358 & 1.7079 & 3.7901 & 6.4966 \\
& 7 & 0 & 0.0508 & 0.0259 & \textcolor{red}{0.0243} & 0.0358 & 0.0708 & 0.1667 \\ 
& & 1 & 0.0508 & 0.0259 & 0.\textcolor{red}{0243} & 0.0358 & 0.0708 & 0.1667 \\ 
& & 2 & 0.7178 & \textcolor{red}{0.5222} & 0.5612 & 0.9985 & 2.6247 & 4.0931 \\
& 8 & 0 & 0.0432 & 0.0187 & \textcolor{red}{0.0150} & 0.0192 & 0.0337 & 0.0724 \\
& & 1 & 0.2184 & 0.1123 & \textcolor{red}{0.0981} & 0.1411 & 0.2818 & 0.6115 \\ 
& & 2 & 0.5925 & 0.3773 & \textcolor{red}{0.3518} & 0.5598 & 1.5068 & 2.3926 \\
& 9 & 0 & 0.0377 & 0.0141 & \textcolor{red}{0.0098} & 0.0111 & 0.0173 & 0.0336 \\
& & 1 & 0.1897 & 0.0847 & \textcolor{red}{0.0642} & 0.0810 & 0.1429 & 0.2982 \\
& & 2 & 0.2184 & 0.1123 & \textcolor{red}{0.0981} & 0.1411 & 0.2818 & 0.6115 \\
& 10 & 0 & 0.0334 & 0.0111 & \textcolor{red}{0.0068} & 0.0068 & 0.0094 & 0.0167 \\
& & 1 & 0.1680 & 0.0661 & \textcolor{red}{0.0441} & 0.0493 & 0.0779 & 0.1510 \\
& & 2 & 0.4459 & 0.2217 & \textcolor{red}{0.1580} & 0.1951 & 0.3551 & 0.7113 \\
\hline
1.5 & 0 & 0 & 0.9038 & 0.8509 & 0.3751 & \textcolor{red}{0.2060} & 0.2134 & 0.9103 \\ 
& & 1 & 2.0309 & 2.6646 & \textcolor{red}{0.3675} & 0.3690 & 0.3926 & 2.1817 \\ 
& & 2 & \textcolor{red}{3.1466} & 5.2795 & 5.9158 & 6.3915 & 9.3943 & 5.84636 \\
& 1 & 0 & 0.7121 & 0.5060 & 0.7230 & \textcolor{red}{0.2875} & 0.6123 & 1.7541 \\
& & 1 & 1.8803 & 1.9898 & 0.5469 & \textcolor{red}{0.2441} & 1.0674 & 1.8833 \\
& & 2 & 3.0244 & 4.3589 & \textcolor{red}{0.7898} & 2.6737 & 2.4258 & 7.3114 \\
& 2 & 0 & \textcolor{red}{0.2476} & 0.2485 & 0.3273 & 0.5260 & 0.8961 & 0.8372 \\
& & 1 & 1.7510 & 1.1513 & 1.810 & 1.2946 & \textcolor{red}{0.9145} & 1.7996 \\
& & 2 & 2.7666 & 2.8912 & 1.0133 & \textcolor{red}{0.3890} & 0.4280 & 3.5673 \\
& 3 & 0 & 0.1414 & \textcolor{red}{0.1185} & 0.1501 & 0.2727 & 0.5895 & 1.0067 \\
& & 1 & 0.7398 & \textcolor{red}{0.6236} & 0.7528 & 1.1439 & 2.3482 & 3.0921 \\
& & 2 & 2.6608 & \textcolor{red}{1.9264} & 2.6575 & 2.4168 & 3.4766 & 4.6030 \\
& 4 & 0 & 0.0961 & \textcolor{red}{0.0627} & 0.0678 & 0.1085 & 0.2218 & 0.4689 \\
& & 1 & 0.4820 & \textcolor{red}{0.3486} & 0.3824 & 0.6340 & 1.6815 & 2.2922 \\
& & 2 & 1.3078 & \textcolor{red}{1.0869} & 1.0912 & 1.6585 & 3.5125 & 4.9771 \\
& 5 & 0 & 0.0723 & \textcolor{red}{0.0372} & 0.0333 & 0.0456 & 0.0827 & 0.1754 \\
& & 1 & 0.3586 & 0.2115 & \textcolor{red}{0.1981} & 0.2956 & 0.6412 & 1.1594 \\
& & 2 & 0.9647 & 0.6623 & \textcolor{red}{0.6284} & 0.9416 & 2.7046 & 3.3749 \\
& 6 & 0 & 0.0581 & 0.0243 & \textcolor{red}{0.0180} & 0.0211 & 0.0335 & 0.0652 \\
& & 1 & 0.2872 & 0.1392 & \textcolor{red}{0.1095} & 0.1408 & 0.2523 & 0.4998 \\
& & 2 & 0.7563 & 0.4416 & \textcolor{red}{0.3599} & 0.4944 & 1.2229 & 1.8582 \\
& 7 & 0 & 0.0486 & 0.0170 & \textcolor{red}{0.0106} & 0.0107 & 0.0149 & 0.0259 \\
& & 1 & 0.2404 & 0.0978 & \textcolor{red}{0.0650} & 0.0723 & 0.1127 & 0.2115 \\
& & 2 & 0.6261 & 0.3134 & \textcolor{red}{0.2165} & 0.2612 & 0.4765 & 0.8950 \\
& 8 & 0 & 0.0419 & 0.0126 & 0.0067 & \textcolor{red}{0.0059} & 0.0072 & 0.0112\\
& & 1 & 0.2074 & 0.0724 & 0.0412 & \textcolor{red}{0.0399} & 0.0549 & 0.0940 \\
& & 2 & 0.5371 & 0.2333 & \textcolor{red}{0.1377} & 0.1453 & 0.2279 & 0.4177 \\
& 9 & 0 & 0.0368 & 0.0097 & 0.0045 & \textcolor{red}{0.0035} & 0.0038 & 0.0052 \\
& & 1 & 0.1826 & 0.0557 & 0.0276 & \textcolor{red}{0.0235} & 0.0288 & 0.0445 \\
& & 2 & 0.4719 & 0.1802 & 0.0922 & \textcolor{red}{0.0856} & 0.1190 & 0.2020 \\
& 10 & 0 & 0.0329 & 0.0077 & 0.0032 & 0.0021 & \textcolor{red}{0.0021} & 0.0026 \\
& & 1 & 0.1633 & 0.0443 & 0.0194 & \textcolor{red}{0.0146} & 0.0161 & 0.0225 \\
& & 2 & 0.4217 & 0.1434 & 0.0645 & \textcolor{red}{0.0531} & 0.0663 & 0.1030 \\
\hline
\end{tabular}
\end{adjustbox}
\caption{Numerical values of $\Delta_k$ for $c_0=0.96$ and $c_{0}=1.5$, multipole number $L$ and overtone $n$. The values highlighted in red color correspond to the minimum value of $\Delta_{k}$}
\label{table:1}
\end{table*}

\begin{table*}[t]
\centering
\begin{adjustbox}{width=11cm,center}
\begin{tabular}{|m{4em}|m{2em}|m{2em}|m{4em}|m{4em}|m{4em}|m{4em}|m{4em}|m{4em}|} 
 \hline
 $c_0$ & $L$ & $n$ &$\Delta_{2}$ & $\Delta_{3}$ &$\Delta_{4}$ &$\Delta_{5}$&$\Delta_{6}$&$\Delta_{7}$ \\  
\hline
3& 0 & 0 & 0.6308 & 0.4292 & 0.9865 & \textcolor{red}{0.4743} & 0.6121 & 0.5187 \\
& & 1 & 2.6174 & 3.3405 & 0.0647 & 0.1875 & \textcolor{red}{0.1864} & 0.0258 \\
& & 2 & 3.9411 & 6.0466 & 6.1787 & \textcolor{red}{5.1435} & 7.8531 & 9.1116 \\
& 1 & 0 & 0.5130 & 0.3770 & 0.2731 & \textcolor{red}{0.1796} & 0.4129 & 0.579402\\
& & 1 & 2.4954 & 2.8692 & \textcolor{red}{0.2091} & 0.2345 & 2.3591 & 2.1394 \\
& & 2 & 3.7755 & 5.5622 & \textcolor{red}{0.4306} & 1.0394 & 7.6925 & 7.7568 \\
& 2 & 0 & 0.2729 & 0.2018 & 0.1641 & 0.1140 & \textcolor{red}{0.0935} & 0.5407 \\
& & 1 & 1.1120 & 0.7857 & 0.4747 & \textcolor{red}{0.2468} & 0.5982 & 1.5273 \\
& & 2 & 3.6226 & 4.3248 & 0.4199 & 0.3637 & \textcolor{red}{0.2410} & 1.7021 \\
& 3 & 0 & 0.1700 & 0.1063 & 0.0881 & 0.0906 & 0.0998 & \textcolor{red}{0.0878} \\
& & 1 & 0.7934 & 0.5037 & 0.3765 & 0.3157 & \textcolor{red}{0.2427} & 0.3020 \\
& & 2 & 1.7864 & 1.1743 & 0.5936 & 0.2717 & \textcolor{red}{0.2088} & 1.2372 \\
& 4 & 0 & 0.1199 & 0.0603 & \textcolor{red}{0.0448} & 0.0466 & 0.0604& 0.0872 \\
& & 1 & 0.5700 & 0.3095 & \textcolor{red}{0.2205} & 0.2208 & 0.2869 & 0.3786 \\
& & 2 & 1.4103 & 0.8412 & 0.5129 & \textcolor{red}{0.4627} & 0.5940 & 0.6875 \\
& 5 & 0 & 0.0919 & 0.0375 & 0.0238 & \textcolor{red}{0.0225} & 0.0281 & 0.0419 \\
& & 1 & 0.4411 & 0.1994 & 0.1250 & \textcolor{red}{0.1213} & 0.16047 & 0.2442 \\
& & 2 & 1.1107 & 0.5749 & 0.3339 & \textcolor{red}{0.3213} & 0.4680 & 0.7215 \\
& 6 & 0 & 0.0745 & 0.0252 & 0.0136 & \textcolor{red}{0.0113} & 0.0129& 0.0181 \\
& & 1 & 0.3602 & 0.1363 & 0.0735 & \textcolor{red}{0.0648} & 0.0798 & 0.1192 \\
& & 2 & 0.9085 & 0.4071 & 0.2080 & \textcolor{red}{0.1898} & 0.2606 & 0.4097 \\
& 7 & 0 & 0.0627 & 0.018 & 0.0083 & \textcolor{red}{0.0061} & 0.0062 & 0.0079 \\
& & 1 & 0.3050 & 0.0983 & 0.0457 & \textcolor{red}{0.0358} & 0.0401 & 0.0557 \\
& & 2 & 0.7701 & 0.2998 & 0.1325 & \textcolor{red}{0.1099} & 0.1380 & 0.2071 \\
& 8 & 0 & 0.0542 & 0.0135 & 0.0054 & 0.0035 & \textcolor{red}{0.0031} & 0.0036 \\
& & 1 & 0.2648 & 0.0740 & 0.0299 & \textcolor{red}{0.0208} & 0.0210 & 0.0267 \\
& & 2 & 0.6702 & 0.2287 & 0.0878 & \textcolor{red}{0.0653} & 0.0745 & 0.1037 \\
& 9 & 0 & 0.0477 & 0.0105 & 0.00375 & 0.0021 & \textcolor{red}{0.0017} & 0.0018 \\
& & 1 & 0.2343 & 0.0577 & 0.0205 & 0.0127 & \textcolor{red}{0.0115} & 0.0134 \\
& & 2 & 0.5944 & 0.1798 & 0.0605 & \textcolor{red}{0.0402} & 0.0417 & 0.0534 \\
& 10 & 0 & 0.0427 & 0.0084 & 0.0026 & 0.0013 & 0.0009 & \textcolor{red}{0.0009} \\
& & 1 & 0.2103 & 0.0462 & 0.0146 & 0.0081 & \textcolor{red}{0.0066} & 0.0070 \\
& & 2 & 0.5347 & 0.1449 & 0.0433 & 0.0258 & \textcolor{red}{0.0243} & 0.0286 \\
\hline
4.86& 0 & 0 & 0.6291 & 0.3608 & \textcolor{red}{0.1609} & 0.2547 & 0.2342 & 0.5114\\
& & 1 & 1.3950 & 0.5022 & \textcolor{red}{0.0403} & 0.1346 & 0.1298 & 0.1342 \\
& & 2 & \textcolor{red}{4.5186} & 6.6421 & 6.6954 & 5.2056 & 7.5131 & 8.7060 \\
& 1 & 0 & 0.5065 & 0.3317 & 0.1831 & \textcolor{red}{0.1774} & 0.2906 & 0.2242 \\
& & 1 & 1.3555 & 0.6158 & \textcolor{red}{0.1262} & 0.1802 & 0.2378 & 0.4410 \\
& & 2 & 4.3407 & 6.3206 & 6.2357 & 7.0961 & \textcolor{red}{2.2726} & 5.2529 \\
& 2 & 0 & 0.2998 & 0.1941 & 0.1208 & \textcolor{red}{0.0547} & 0.1275 & 0.3779 \\
& & 1 & 1.1617 & 0.6579 & 0.2795 & \textcolor{red}{0.2398} & 0.5222 & 0.8681 \\
& & 2 & 4.2160 & 5.2480 & 0.3368 & 0.3719 & \textcolor{red}{0.3337} & 0.9302 \\
& 3 & 0 & 0.1934 & 0.1083 & 0.0724 & 0.0523 & \textcolor{red}{0.0308} & 0.0622 \\
& & 1 & 0.8688 & 0.4819 & 0.2690 & 0.1290 & \textcolor{red}{0.1237} & 0.4982 \\
& & 2 & 1.9094 & 1.0190 & 0.3193 & \textcolor{red}{0.1109} & 0.3859 & 1.1002 \\
& 4 & 0 & 0.1383 & 0.0637 & 0.0399 & 0.0325 & 0.0305 & \textcolor{red}{0.0268} \\
& & 1 & 0.6445 & 0.3140 & 0.1775 & 0.1270 & 0.0988 & \textcolor{red}{0.0602} \\
& & 2 & 1.5445 & 0.8018 & 0.3405 & 0.1714 & \textcolor{red}{0.1029} & 0.3035 \\
& 5 & 0 & 0.1066 & 0.0405 & 0.0223 & 0.0174 & \textcolor{red}{0.0174} & 0.0198 \\
& & 1 & 0.5052 & 0.2087 & 0.1079 & 0.0821 & \textcolor{red}{0.0814} & 0.0878 \\
& & 2 & 1.2444 & 0.5794 & 0.2526 & 0.1760 & 0.1773 & \textcolor{red}{0.1619} \\
& 6 & 0 & 0.0865 & 0.0275 & 0.0131 & 0.0093 & \textcolor{red}{0.0088} & 0.0101 \\
& & 1 & 0.4149 & 0.1452 & 0.0662 & 0.0479 & \textcolor{red}{0.0478} & 0.0568 \\
& & 2 & 1.0310 & 0.4221 & 0.1689 & \textcolor{red}{0.1213} & 0.1314 & 0.1576 \\
& 7 & 0 & 0.0729 & 0.0198 & 0.0082 & 0.0052 & \textcolor{red}{0.0045} & 0.0049 \\
& & 1 & 0.3523 & 0.1058 & 0.0423 & 0.0279 & \textcolor{red}{0.0262} & 0.0304 \\
& & 2 & 0.8801 & 0.3159 & 0.1123 & \textcolor{red}{0.0764} & 0.0800 & 0.0979 \\
& 8 & 0 & 0.0630 & 0.0149 & 0.0054 & 0.0030 & 0.0024 & \textcolor{red}{0.0024} \\
& & 1 & 0.3065 & 0.0801 & 0.0282 & 0.0168 & \textcolor{red}{0.0145} & 0.0158 \\
& & 2 & 0.7690 & 0.2434 & 0.0764 & 0.0478 & \textcolor{red}{0.0468} & 0.0548 \\
& 9 & 0 & 0.0555 & 0.0116 & 0.0037 & 0.0018 & 0.0013 & \textcolor{red}{0.0012} \\
& & 1 & 0.2714 & 0.0627 & 0.0195 & 0.0105 & \textcolor{red}{0.0083} & 0.0083 \\
& & 2 & 0.6838 & 0.1925 & 0.0536 & 0.0305 & \textcolor{red}{0.0275} & 0.0302 \\
& 10 & 0 & 0.0497 & 0.0093 & 0.0027 & 0.0012 & 0.0007 & \textcolor{red}{0.0006} \\
& & 1 & 0.2437 & 0.0504 & 0.0141 & 0.0068 & 0.0049 & \textcolor{red}{0.0045} \\
& & 2 & 0.6162 & 0.1558 & 0.0388 & 0.0200 & \textcolor{red}{0.0166} & 0.0169 \\
\hline
\end{tabular}
\end{adjustbox}
\caption{Numerical values of $\Delta_k$ for $c_0=3$ and $c_{0}=4.86$, multipole number $L$ and overtone $n$. The values highlighted in red color correspond to the minimum value of $\Delta_{k}$.}
\label{table:2}
\end{table*}

\begin{table*}[t]
\centering
\begin{adjustbox}{width=10cm,center}
\begin{tabular}{|m{3em}|m{2em}|m{2em}|m{3em}|m{3em}|m{3em}|m{3em}|m{3em}|m{3em}|m{3em}|} 
 \hline
 $c_0$ & $L$ & $n$ & $\omega_{1}$&$\omega_{2}$ & $\omega_{3}$ &$\omega_{4}$ &$\omega_{5}$&$\omega_{6}$&$\omega_{7}$ \\  
\hline
0.96& 0 & 0  & -0.514 & -1.123 & \hspace{10pt}-& \hspace{10pt}-& -0.088 & -1.031 & \hspace{10pt}-\\
&& 1 & -0.995 & -2.638 & \hspace{10pt}-& \hspace{10pt}-& -2.219 & -1.588 & \hspace{10pt}-\\
&& 2 & -1.315 & -4.270 & \hspace{10pt}-& \hspace{10pt}-& -8.537 & -2.579 & \hspace{10pt}-\\
& 1 & 0 & -0.459 & -0.844 & \hspace{10pt}-& \hspace{10pt}-& -0.663 & -0.944 & -1.970 \\
&& 1 & -1.011 & -2.329 & \hspace{10pt}-& \hspace{10pt}-& \hspace{10pt}-& \hspace{10pt}-& \hspace{10pt}-\\
&& 2 & -1.376 & -3.841 & \hspace{10pt}-& \hspace{10pt}-& \hspace{10pt}-& \hspace{10pt}-& \hspace{10pt}-\\
& 2 & 0 &-0.416 & -0.557 & \hspace{10pt}-& \hspace{10pt}-& -0.936 & -2.326 & \hspace{10pt}-\\
&& 1 & -1.028 & -1.964 & \hspace{10pt}-& \hspace{10pt}-& -1.616 & -3.821 & \hspace{10pt}-\\
&& 2 & -1.456 & -3.389 & \hspace{10pt}-& \hspace{10pt}-& \hspace{10pt}-& \hspace{10pt}-& \hspace{10pt}-\\
& 3 & 0 & -0.394 & -0.452 & -0.204 & -0.167 & -0.727 & -1.698 & \hspace{10pt}-\\
&& 1 & -1.043 & -1.655 & \hspace{10pt}-& \hspace{10pt}-& -1.919 & -4.908 & \hspace{10pt}-\\
&& 2 & -1.526 & -3.028 & \hspace{10pt}-& \hspace{10pt}-& -2.381 & -6.706 & \hspace{10pt}-\\
& 4 & 0 & -0.382 & -0.413 & -0.277 & -0.257 & -0.530 & -0.756 & \hspace{10pt}-\\
&& 1 & -1.052 & -1.442 & -0.699 & -0.383 & -1.656 & -3.816 & \hspace{10pt}-\\
&& 2 & -1.583 & -2.737 & \hspace{10pt}-& \hspace{10pt}-& -2.622 & -7.006 & \hspace{10pt}-\\
& 5 & 0 & -0.375 & -0.394 & -0.309 & -0.299 & -0.431 & -0.482 & \hspace{10pt}-\\
&& 1 & -1.059 & -1.318 & -0.842 & -0.672 & -1.396 & -2.513 & \hspace{10pt}-\\
&& 2 & -1.626 & -2.507 & -1.531 & -0.306 & -2.361 & -5.555 & \hspace{10pt}-\\
& 6 & 0 & -0.371 & -0.384 & -0.327 & -0.322 & -0.389 & -0.405 & -0.100 \\
&& 1 & -1.063 & -1.245 & -0.919 & -0.827 & -1.233 & -1.645 & \hspace{10pt}-\\
&& 2 & -1.659 & -2.333 & -1.413 & -0.857 & -2.115 & -4.085 & \hspace{10pt}-\\
& 7 & 0 & -0.368 & -0.378 & -0.337 & -0.334 & -0.371 & -0.377 & -0.248 \\
&& 1 & -1.067 & -1.201 & -0.964 & -0.912 & -1.146 & -1.299 & -0.135 \\
&& 2 & -1.684 & -2.207 & -1.491 & -1.185 & -1.950 & -2.963 & \hspace{10pt}-\\
& 8 & 0 & -0.366 & -0.373 & -0.342 & -0.341 & -0.362 & -0.365 & -0.306 \\
&& 1 & -1.069 & -1.173 & -0.993 & -0.962 & -1.102 & -1.168 & -0.662 \\
&& 2 & -1.704 & -2.117 & -1.558 & -1.376 & -1.852 & -2.319 & \hspace{10pt}-\\
& 9 & 0 & -0.365 & -0.371 & -0.346 & -0.345 & -0.359 & -0.360 & -0.330 \\
&& 1 & -1.071 & -1.153 & -1.012 & -0.992 & -1.081 & -1.113 & -0.863 \\
&& 2 & -1.718 & -2.052 & -1.607 & -1.493 & -1.799 & -2.022 & -0.824 \\
& 10 & 0 & -0.364 & -0.369 & -0.349 & -0.348 & -0.357 & -0.357 & -0.342 \\
&& 1 & -1.072 & -1.139 & -1.026 & -1.012 & -1.071 & -1.087 & -0.955 \\
&& 2 & -1.731 & -2.005 & -1.643 & -1.568 & -1.771 & -1.887 & -1.280 \\
\hline
1.5 & 0 & 0 & -0.594 & -1.163 & \hspace{10pt}-& \hspace{10pt}-& \hspace{10pt}-& \hspace{10pt}-& \hspace{10pt}-\\
&& 1 & -1.179 & -2.800 & \hspace{10pt}-& \hspace{10pt}-& \hspace{10pt}-& \hspace{10pt}-& \hspace{10pt}-\\
&& 2 & -1.565 & -4.539 & \hspace{10pt}-& \hspace{10pt}-& -6.909 & -2.289 & \hspace{10pt}-\\
& 1 & 0 & -0.536 & -0.857 & \hspace{10pt}-& \hspace{10pt}-& -0.491 & -0.511 & -1.543 \\
&& 1 & -1.201 & -2.471 & \hspace{10pt}-& \hspace{10pt}-& \hspace{10pt}-& \hspace{10pt}-& -0.719 \\
&& 2 & -1.644 & -4.096 & \hspace{10pt}-& \hspace{10pt}-& \hspace{10pt}-& \hspace{10pt}-& \hspace{10pt}-\\
& 2 & 0 & -0.491 & -0.608 & -0.262 & -0.199 & -0.665 & -1.086 & \hspace{10pt}-\\
&& 1 & -1.226 & -2.095 & \hspace{10pt}-& \hspace{10pt}-& -1.331 & -2.382 & -2.123 \\
&& 2 & -1.744 & -3.636 & \hspace{10pt}-& \hspace{10pt}-& \hspace{10pt}-& \hspace{10pt}-& \hspace{10pt}-\\
& 3 & 0 & -0.467 & -0.519 & -0.353 & -0.325 & -0.558 & -0.699 & \hspace{10pt}-\\
&& 1 & -1.246 & -1.799 & -1.132 & -0.582 & -1.558 & -2.782 & \hspace{10pt}-\\
&& 2 & -1.832 & -3.271 & \hspace{10pt}-& \hspace{10pt}-& -1.920 & -4.213 & \hspace{10pt}-\\
& 4 & 0 & -0.455 & -0.483 & -0.389 & -0.377 & -0.482 & -0.517 & -0.105 \\
&& 1 & -1.260 & -1.613 & -1.144 & -0.937 & -1.461 & -2.066 & \hspace{10pt}-\\
&& 2 & -1.901 & -2.988 & -2.224 & -0.815 & -2.241 & -4.099 & \hspace{10pt}-\\
& 5 & 0 & -0.448 & -0.466 & -0.406 & -0.401 & -0.450 & -0.460 & -0.313 \\
&& 1 & -1.269 & -1.508 & -1.192 & -1.097 & -1.371 & -1.586 & -0.448 \\
&& 2 & -1.954 & -2.775 & -2.012 & -1.451 & -2.226 & -3.259 & \hspace{10pt}-\\
& 6 & 0 & -0.444 & -0.456 & -0.415 & -0.412 & -0.437 & -0.441 & -0.383 \\
&& 1 & -1.275 & -1.446 & -1.223 & -1.175 & -1.322 & -1.402 & -0.971 \\
&& 2 & -1.994 & -2.622 & -2.009 & -1.739 & -2.180 & -2.648 & -0.820 \\
& 7 & 0 & -0.441 & -0.450 & -0.420 & -0.418 & -0.432 & -0.434 & -0.409 \\
&& 1 & -1.280 & -1.408 & -1.243 & -1.216 & -1.300 & -1.333 & -1.146 \\
&& 2 & -2.024 & -2.515 & -2.035 & -1.888 & -2.148 & -2.357 & -1.602 \\
& 8 & 0 & -0.439 & -0.446 & -0.423 & -0.422 & -0.430 & -0.431 & -0.419 \\
&& 1 & -1.283 & -1.382 & -1.256 & -1.240 & -1.290 & -1.306 & -1.216 \\
&& 2 & -2.048 & -2.439 & -2.059 & -1.972 & -2.132 & -2.229 & -1.869 \\
& 9 & 0& -0.438 & -0.443 & -0.425 & -0.425 & -0.430 & -0.430 & -0.424 \\
&& 1 & -1.285 & -1.365 & -1.265 & -1.255 & -1.286 & -1.294 & -1.249 \\
&& 2 & -2.066 & -2.384 & -2.078 & -2.023 & -2.126 & -2.175 & -1.989 \\
& 10 & 0 & -0.437 & -0.441 & -0.427 & -0.426 & -0.429 & -0.430 & -0.427 \\
&& 1 & -1.287 & -1.352 & -1.271 & -1.265 & -1.285 & -1.289 & -1.265 \\
&& 2 & -2.080 & -2.344 & -2.092 & -2.057 & -2.125 & -2.151 & -2.050 \\
\hline
\end{tabular}
\end{adjustbox}
\caption{Numerical values of $Im(\omega_k)$ at each order $k$ for $c_{0}=0.96$ and $c_{0}=1.5$. Entries with a dash indicate that the approach does not lead to a suitable value.}
\label{table:3}
\end{table*}

\begin{table*}[t]
\centering
\begin{adjustbox}{width=10cm,center}
\begin{tabular}{|m{3em}|m{2em}|m{2em}|m{3em}|m{3em}|m{3em}|m{3em}|m{3em}|m{3em}|m{3em}|} 
 \hline
 $c_0$ & $L$ & $n$ & $\omega_{1}$&$\omega_{2}$ & $\omega_{3}$ &$\omega_{4}$ &$\omega_{5}$&$\omega_{6}$&$\omega_{7}$ \\  
\hline
3& 0 & 0 & -0.800 & -1.394 & -1.187 & -0.793 & \hspace{10pt}-& \hspace{10pt}-& -0.307 \\
&& 1 & -1.617 & -3.439 & \hspace{10pt}-& \hspace{10pt}-& \hspace{10pt}-& \hspace{10pt}-& \hspace{10pt}-\\
&& 2 & -2.156 & -5.579 & \hspace{10pt}-& \hspace{10pt}-& -6.751 & -1.998 & \hspace{10pt}-\\
& 1 & 0 & -0.732 & -1.088 & -0.598 & -0.334 & -0.450 & -0.351 & -1.015 \\
&& 1 & -1.636 & -3.120 & \hspace{10pt}-& \hspace{10pt}-& \hspace{10pt}-& \hspace{10pt}-& -2.167 \\
&& 2 & -2.237 & -5.160 & \hspace{10pt}-& \hspace{10pt}-& \hspace{10pt}-& \hspace{10pt}-& -9.043 \\
& 2 & 0 & -0.672 & -0.821 & -0.507 & -0.440 & -0.628 & -0.662 & -0.810 \\
&& 1 & -1.659 & -2.718 & -2.237 & -1.342 & -1.586 & -1.416 & -2.247 \\
&& 2 & -2.348 & -4.670 & \hspace{10pt}-& \hspace{10pt}-& \hspace{10pt}-& \hspace{10pt}-& \hspace{10pt}-\\
& 3 & 0 & -0.637 & -0.707 & -0.538 & -0.512 & -0.628 & -0.660 & -0.513 \\
&& 1 & -1.679 & -2.382 & -1.792 & -1.383 & -1.762 & -2.013 & -1.824 \\
&& 2 & -2.452 & -4.253 & -3.694 & -2.550 & -2.683 & -2.944 & -3.046 \\
& 4 & 0 & -0.617 & -0.657 & -0.556 & -0.546 & -0.605 & -0.617 & -0.520 \\
&& 1 & -1.693 & -2.159 & -1.709 & -1.540 & -1.789 & -1.958 & -1.561 \\
&& 2 & -2.538 & -3.922 & -3.168 & -2.403 & -2.810 & -3.326 & -2.880 \\
& 5 & 0 & -0.606 & -0.631 & -0.566 & -0.561 & -0.591 & -0.595 & -0.550 \\
&& 1 & -1.703 & -2.025 & -1.705 & -1.626 & -1.770 & -1.847 & -1.606 \\
&& 2 & -2.606 & -3.670 & -2.953 & -2.573 & -2.883 & -3.209 & -2.672 \\
& 6 & 0 & -0.598 & -0.616 & -0.571 & -0.568 & -0.584 & -0.586 & -0.566 \\
&& 1 & -1.710 & -1.943 & -1.712 & -1.671 & -1.753 & -1.787 & -1.666 \\
&& 2 & -2.658 & -3.485 & -2.890 & -2.692 & -2.894 & -3.060 & -2.712 \\
& 7 & 0 & -0.594 & -0.606 & -0.573 & -0.572 & -0.581 & -0.582 & -0.572\\
&& 1 & -1.716 & -1.891 & -1.718 & -1.695 & -1.743 & -1.759 & -1.699 \\
&& 2 & -2.699 & -3.352 & -2.873 & -2.762 & -2.890 & -2.972 & -2.779 \\
& 8 & 0 & -0.590 & -0.600 & -0.575 & -0.574 & -0.579 & -0.580 & -0.575 \\
&& 1 & -1.719 & -1.856 & -1.722 & -1.708 & -1.738 & -1.746 & -1.715 \\
&& 2 & -2.731 & -3.256 & -2.871 & -2.804 & -2.886 & -2.928 & -2.823 \\
& 9 & 0 & -0.588 & -0.596 & -0.576 & -0.575 & -0.579 & -0.579 & -0.577 \\
&& 1 & -1.722 & -1.832 & -1.725 & -1.717 & -1.736 & -1.740 & -1.723 \\
&& 2 & -2.756 & -3.186 & -2.872 & -2.830 & -2.884 & -2.906 & -2.848 \\
& 10 & 0 & -0.586 & -0.593 & -0.576 & -0.576 & -0.578 & -0.578 & -0.577 \\
&& 1 & -1.724 & -1.814 & -1.727 & -1.722 & -1.734 & -1.737 & -1.728 \\
&& 2 & -2.772 & -3.134 & -2.874 & -2.846 & -2.883 & -2.895 & -2.862 \\
\hline
4.86 & 0 & 0 & -0.928 & -1.546 & -1.285 & -1.049 & -0.991 & -0.542 & -0.563 \\
&& 1 & -1.893 & -3.855 & -3.695 & -3.653 & -3.652 & -3.392 & -3.392 \\
&& 2 & -2.529 & -6.253 & \hspace{10pt}-& \hspace{10pt}-& -7.172 & -2.505 & \hspace{10pt}-\\
& 1 & 0 & -0.856 & -1.244 & -0.793 & -0.591 & -0.569 & -0.443 & -0.859 \\
&& 1 & -1.907 & -3.541 & -3.270 & -3.124 & -3.117 & -2.831 & -2.860 \\
&& 2 & -2.607 & -5.846 & \hspace{10pt}-& \hspace{10pt}-& -6.747 & -7.421 & -8.604 \\
& 2 & 0 & -0.786 & -0.961 & -0.644 & -0.581 & -0.683 & -0.675 & -0.913 \\
&& 1 & -1.927 & -3.124 & -2.630 & -2.110 & -2.114 & -1.706 & -2.345 \\
&& 2 & -2.719 & -5.339 & \hspace{10pt}-& \hspace{10pt}-& \hspace{10pt}-& \hspace{10pt}-& \hspace{10pt}-\\
& 3 & 0 & -0.743 & -0.827 & -0.647 & -0.622 & -0.704 & -0.714 & -0.707 \\
&& 1 & -1.944 & -2.761 & -2.177 & -1.837 & -2.029 & -2.073 & -2.254 \\
&& 2 & -2.829 & -4.889 & -4.293 & -3.676 & -3.655 & -3.464 & -3.689 \\
& 4 & 0 & -0.718 & -0.766 & -0.656 & -0.645 & -0.692 & -0.700 & -0.658 \\
&& 1 & -1.958 & -2.508 & -2.042 & -1.884 & -2.053 & -2.136 & -2.037 \\
&& 2 & -2.922 & -4.524 & -3.764 & -3.207 & -3.366 & -3.533 & -3.522 \\
& 5 & 0 & -0.703 & -0.733 & -0.661 & -0.656 & -0.682 & -0.685 & -0.660 \\
&& 1 & -1.967 & -2.351 & -2.011 & -1.934 & -2.042 & -2.091 & -1.983 \\
&& 2 & -2.997 & -4.240 & -3.509 & -3.187 & -3.371 & -3.539 & -3.362 \\
& 6 & 0 & -0.693 & -0.714 & -0.664 & -0.661 & -0.675 & -0.677 & -0.664 \\
&& 1 & -1.974 & -2.253 & -2.003 & -1.962 & -2.029 & -2.052 & -1.986 \\
&& 2 & -3.056 & -4.029 & -3.409 & -3.231 & -3.369 & -3.471 & -3.313 \\
& 7 & 0 & -0.687 & -0.702 & -0.665 & -0.664 & -0.672 & -0.673 & -0.666 \\
&& 1 & -1.979 & -2.190 & -2.001 & -1.978 & -2.019 & -2.030 & -1.994 \\
&& 2 & -3.103 & -3.876 & -3.370 & -3.266 & -3.361 & -3.416 & -3.313 \\
& 8 & 0 & -0.682 & -0.694 & -0.666 & -0.665 & -0.670 & -0.670 & -0.667 \\
&& 1 & -1.982 & -2.147 & -2.001 & -1.987 & -2.012 & -2.018 & -1.998 \\
&& 2 & -3.139 & -3.764 & -3.352 & -3.289 & -3.352 & -3.382 & -3.321 \\
& 9 & 0 & -0.679 & -0.689 & -0.666 & -0.666 & -0.669 & -0.669 & -0.667 \\
&& 1 & -1.985 & -2.117 & -2.001 & -1.992 & -2.008 & -2.011 & -2.000 \\
&& 2 & -3.169 & -3.682 & -3.344 & -3.304 & -3.346 & -3.363 & -3.327 \\
& 10 & 0 & -0.677 & -0.684 & -0.666 & -0.666 & -0.668 & -0.668 & -0.667 \\
&& 1 & -1.987 & -2.095 & -2.000 & -1.995 & -2.006 & -2.007 & -2.001 \\
&& 2 & -3.192 & -3.620 & -3.340 & -3.313 & -3.342 & -3.351 & -3.330 \\
\hline
\end{tabular}
\end{adjustbox}
\caption{Numerical values of $Im(\omega_k)$ at each order $k$ for $c_{0}=3$ and $c_{0}=4.86$. Entries with a dash indicate that the approach does not lead to a suitable value.}
\label{table:4}
\end{table*}

To complement the analysis, we plot both the real and the imaginary part of the QNM frequencies as a function of $c_{0}$ by setting different values of $L$ and varying the overtone $n$. Furthermore, based on the above results we will perform the analysis for values of $L$ we are confident of with our approach, namely, $L=5,6,...,10$.

\begin{figure*}[hbt!]
\centering
\includegraphics[width=0.4\textwidth]{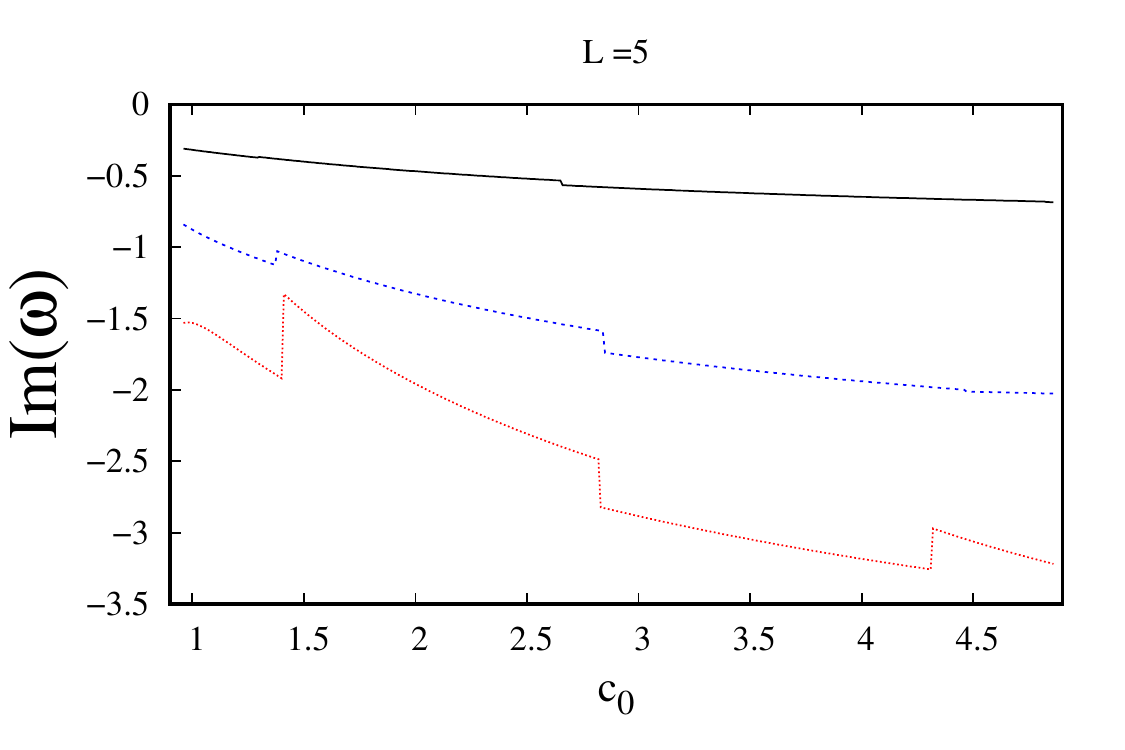}  \
\includegraphics[width=0.4\textwidth]{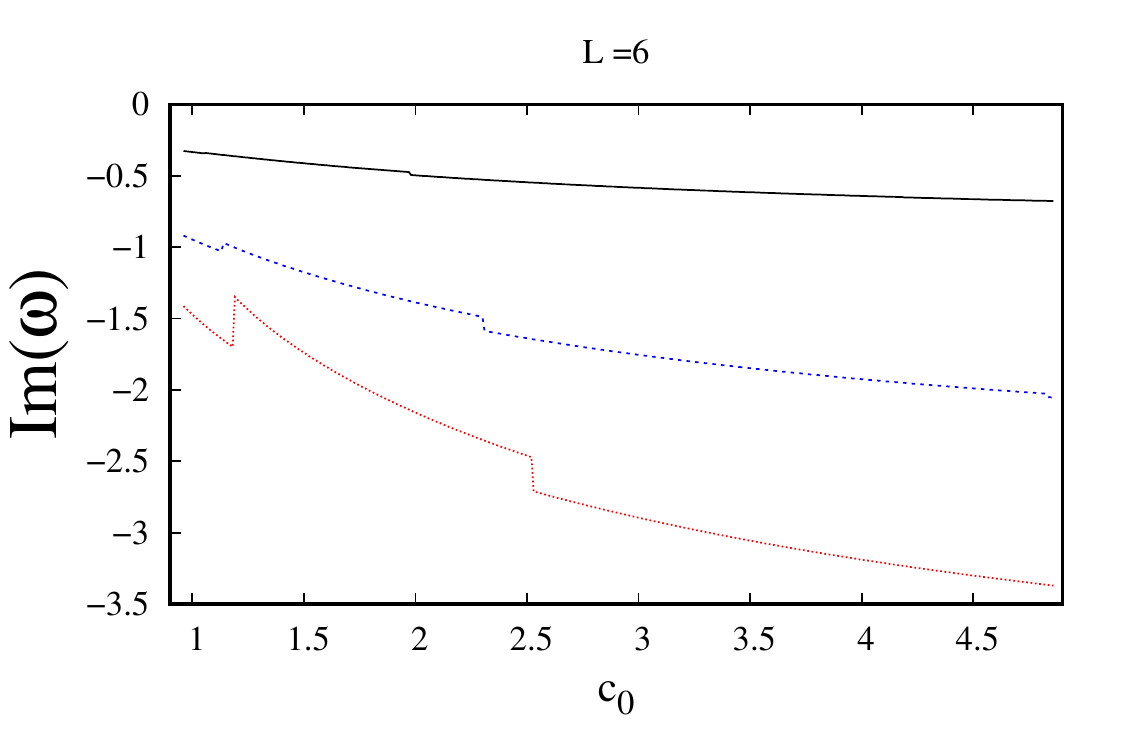}  \
\includegraphics[width=0.4\textwidth]{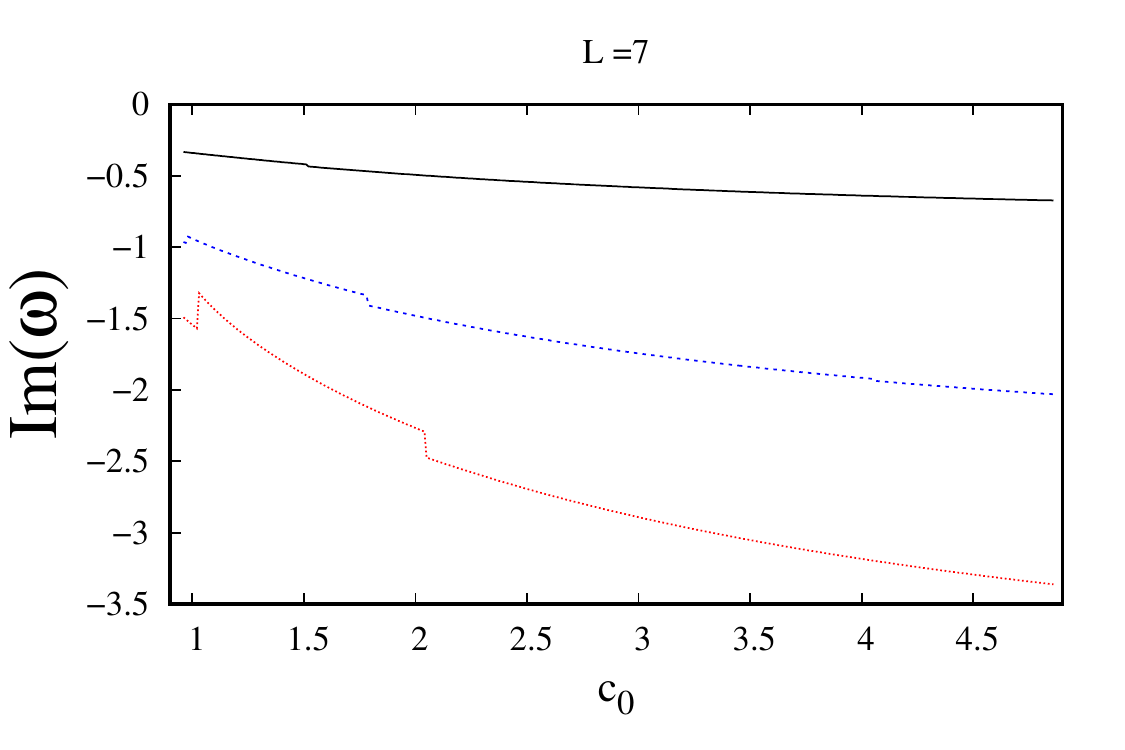}  \
\includegraphics[width=0.4\textwidth]{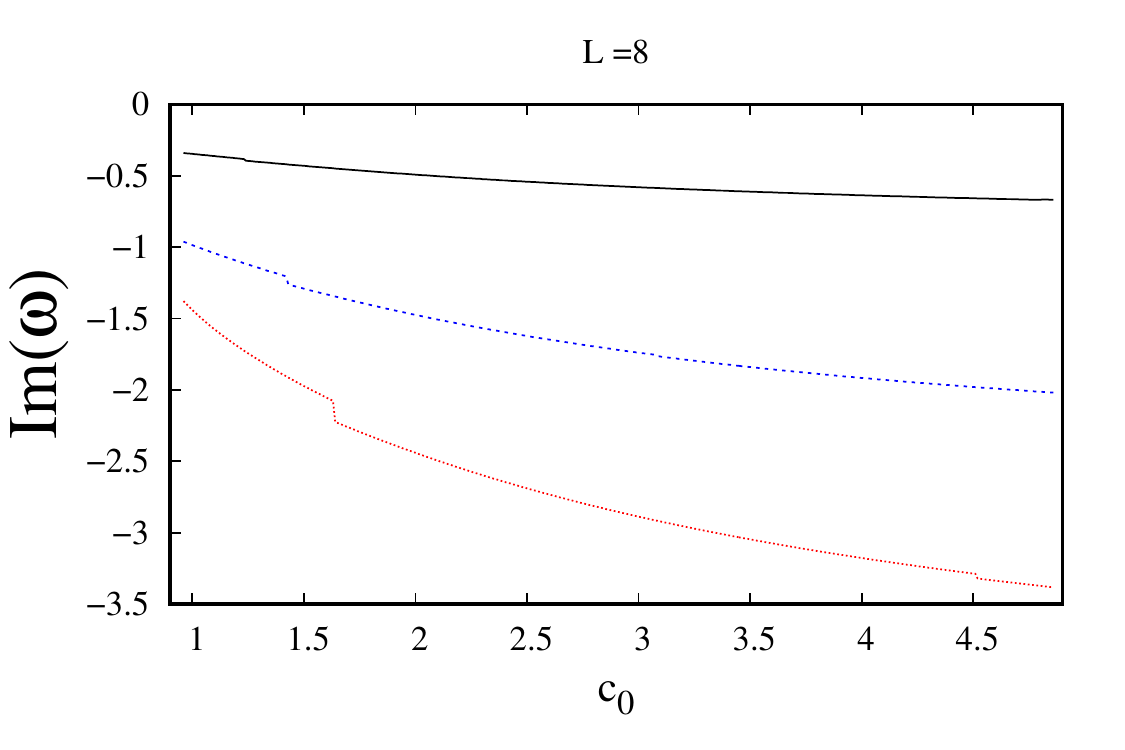}  \
\includegraphics[width=0.4\textwidth]{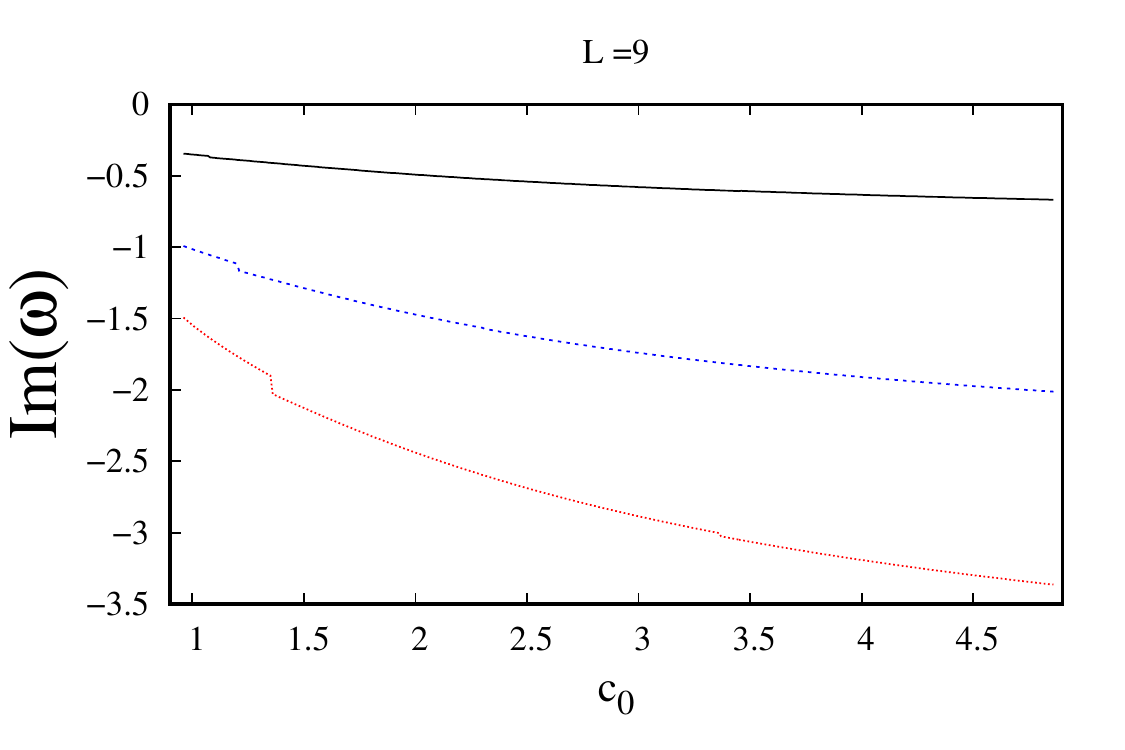}
\includegraphics[width=0.4\textwidth]{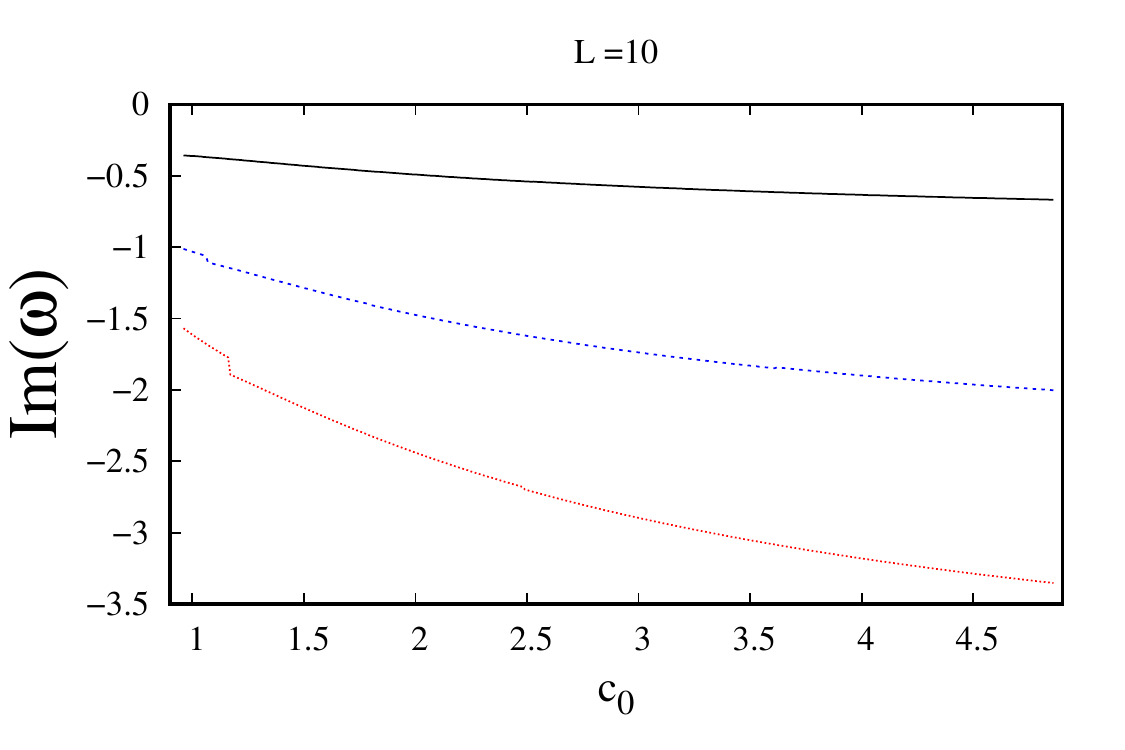}  \
\caption{\label{imaginary}
Imaginary part of the QNM frequencies as a function of the free parameter $c_0$ for $L=5,6,...,10$ and $n=0$ (black line), $n=1$ (blue line), $n=2$ (red line).}
\end{figure*}

In Fig. \ref{imaginary} we show the $Im(\omega)$
as a function of $c_{0}$ for different values of $L$ and $n$. It is worth emphasizing that, the plots are not parameterized with the same $k$ but each point corresponds to the best WKB order. To be more precise, in order to obtain Fig. \ref{imaginary} we have identified the best order $k$ from
the minimum value of $\Delta_{k}$ from a table (not shown here) generated by setting $c_{0}$ in a step of $0.01$. Higher precision is possible but the computational time increases considerably. Finally, we have computed $Im(\omega)$ for each $k$ (associated to the minimum $\Delta_{k}$) so, as said before,
not all the points in Fig. \ref{imaginary} corresponds to the same order. It is worth noticing that $|Im(\omega)|$ increases as $c_{0}$ increases and as a consequence, the damping factor grows with the free parameter $c_{0}$. Even more, for $n=0$ we observe that the damping factor is around zero for some velues of $c_{0}$ mimicking a well--known effect observed for massive scalar fields in the Reissner-Nordstr\"{o}m background, namely, the so--called quasi-resonance for the fundamental mode (see \cite{Churilova:2019qph}, for example). 

\begin{figure*}[hbt!]
\centering
\includegraphics[width=0.4\textwidth]{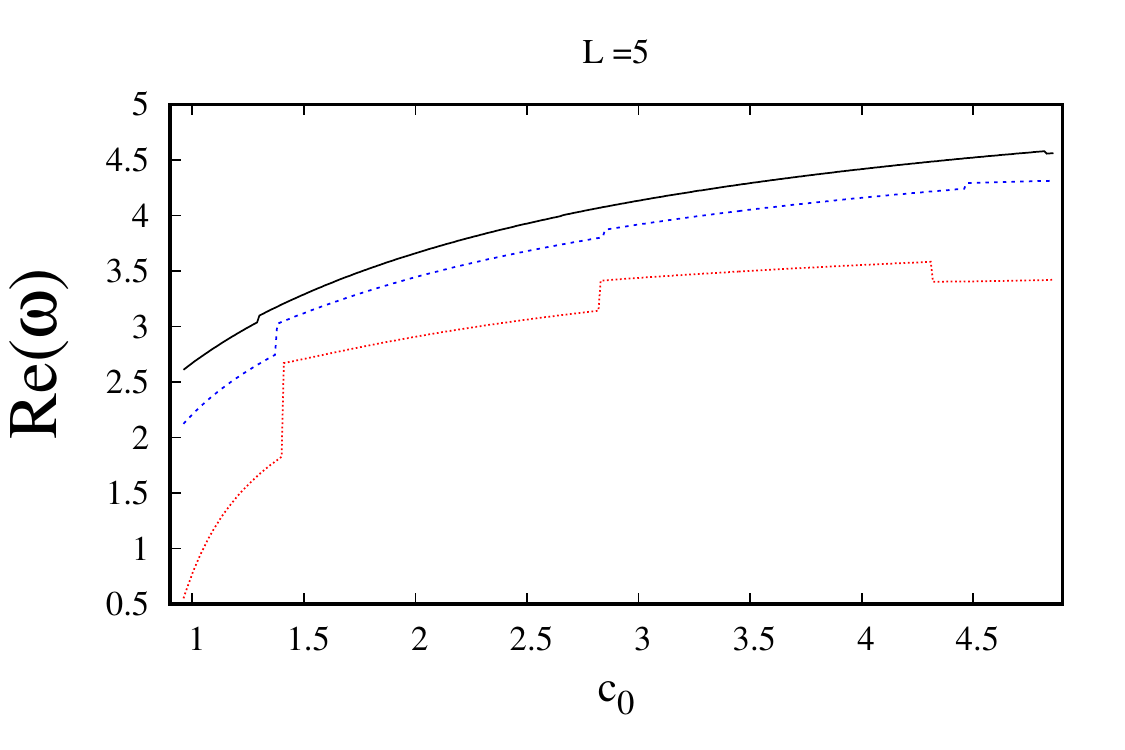}  \
\includegraphics[width=0.4\textwidth]{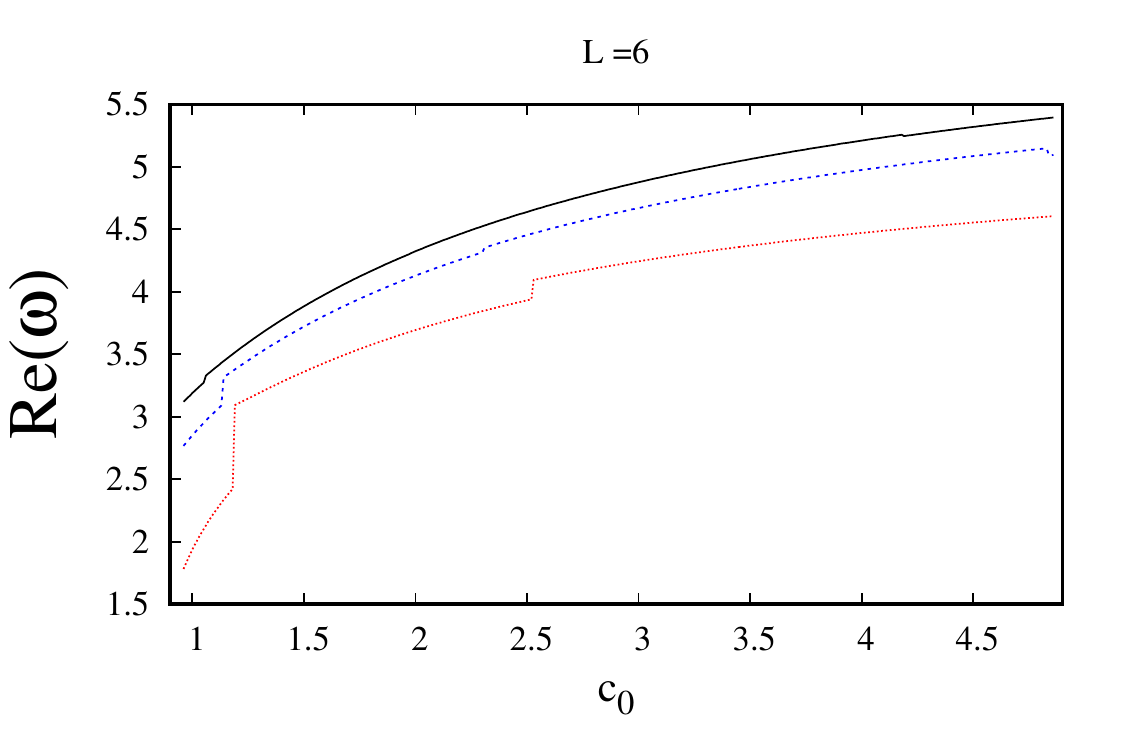}  \
\includegraphics[width=0.4\textwidth]{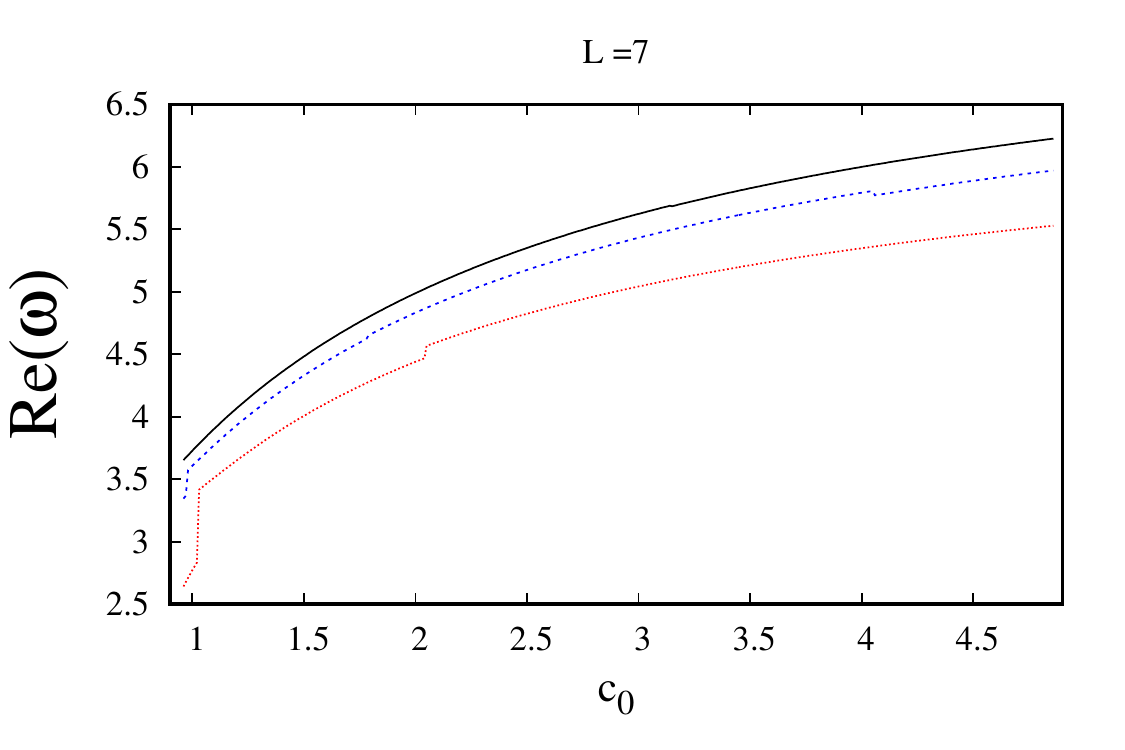}  \
\includegraphics[width=0.4\textwidth]{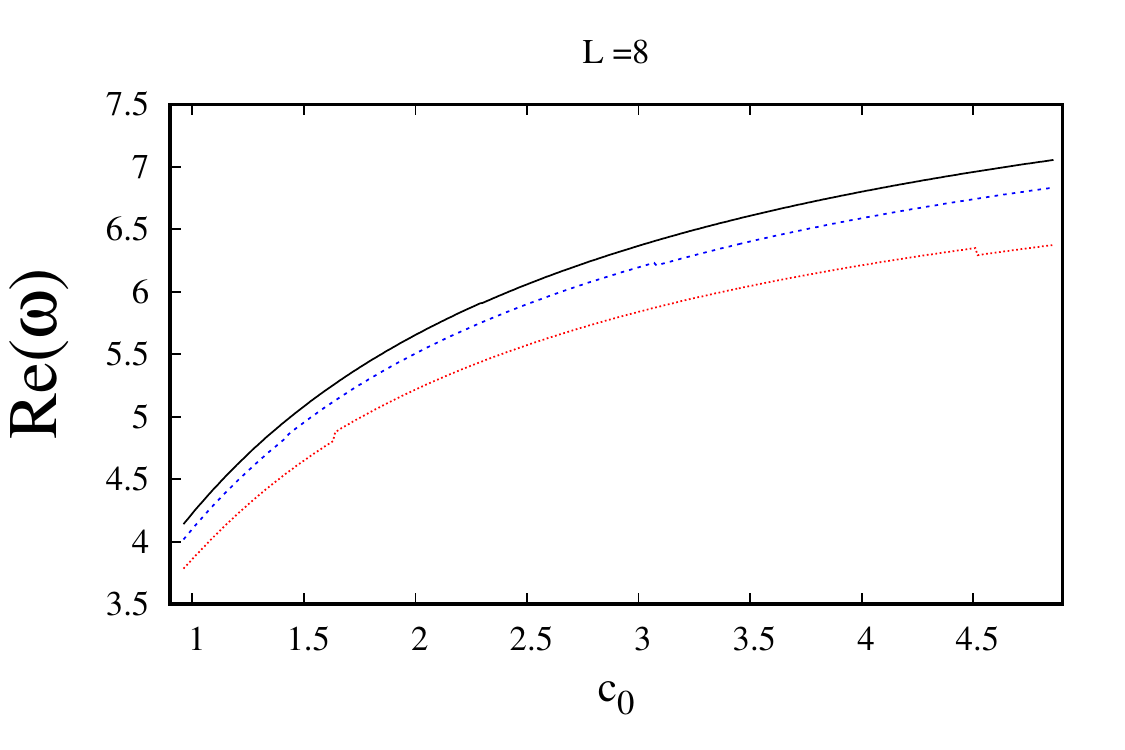}  \
\includegraphics[width=0.4\textwidth]{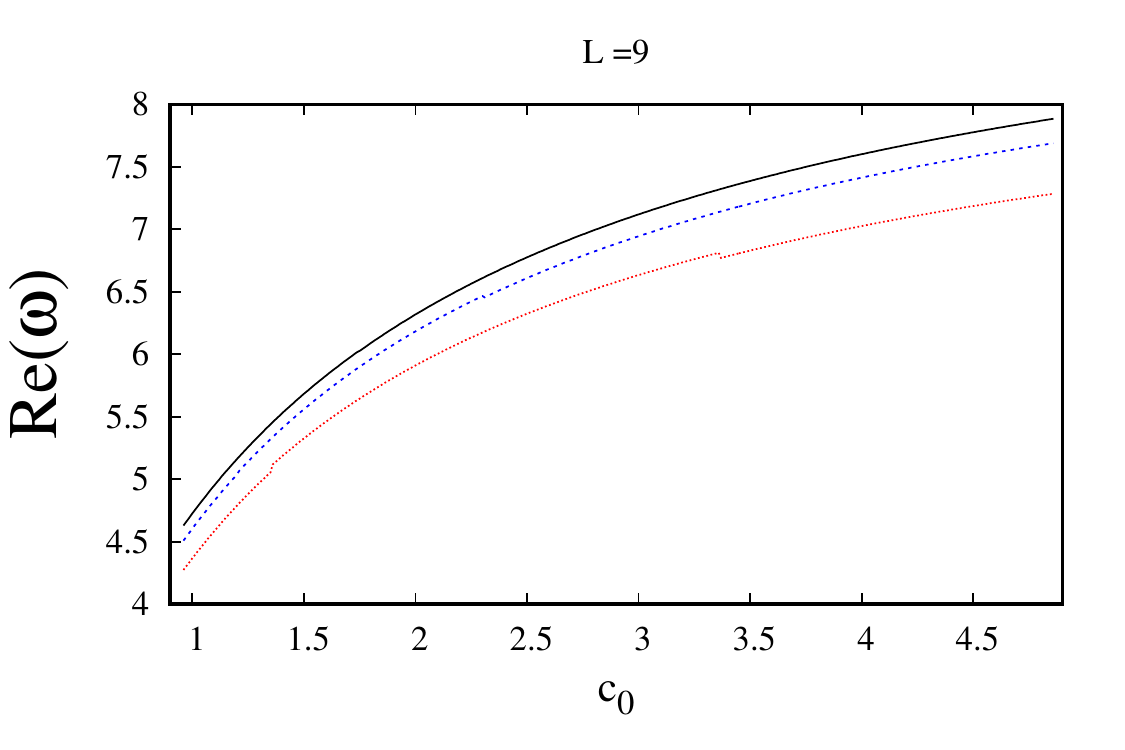}
\includegraphics[width=0.4\textwidth]{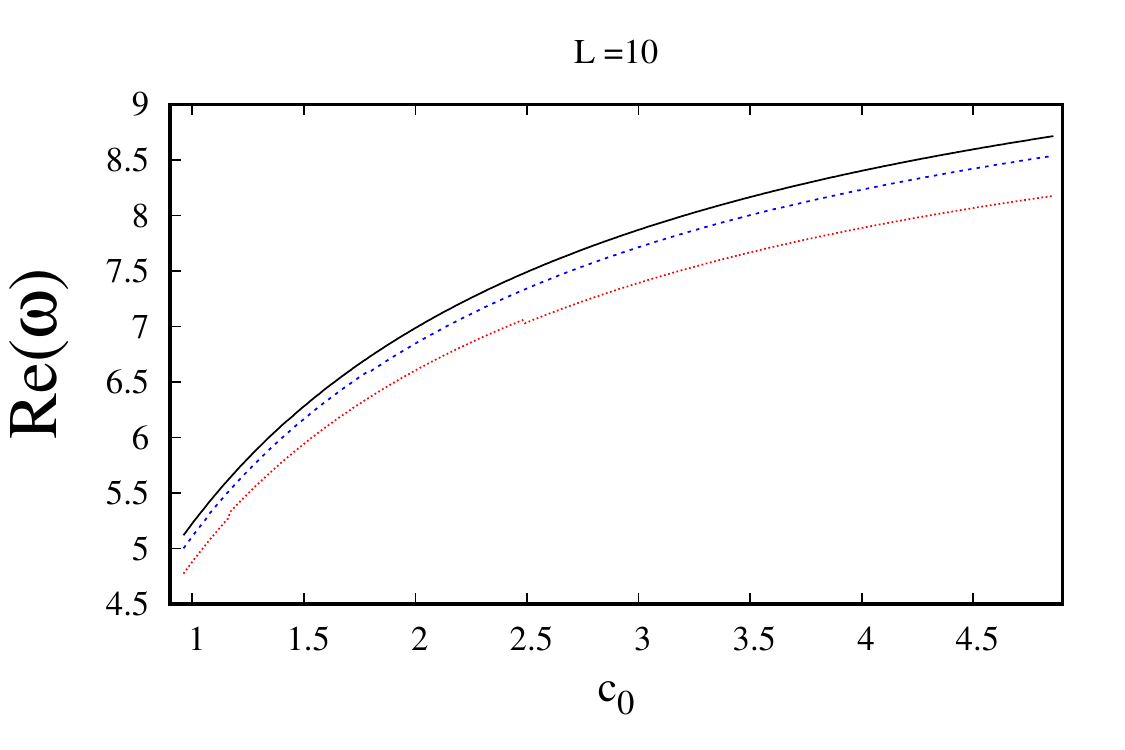}  \
\caption{\label{real}
Real part of the QNM frequency as a function of $c_0$ for for $L=5,6,...,10$ and $n=0$ (black line), $n=1$ (blue line), $n=2$ (red line). }
\end{figure*}

In Fig. \ref{real} we show the $Re(\omega)$ as a function of $c_{0}$ for different values of $L$ and $n$ following the same procedure described previously for the computation of $Im(\omega)$. In this case, we observe that the frequency of the signal increases with $c_{0}$.

\section{Conclusion}
In this work, we obtained the frequencies of the QNM at different WKB-orders for a model of a traversable wormhole with Casimir--like complexity. All the results are shown as a function of the free parameter $c_0$ of this wormhole which is restricted to the values  $0.950679 < c_0 < 4.86215$. It was shown that the effective potential is bell--shaped as a function of the tortoise coordinate. We estimated the error at each order and use this information to plot both the imaginary and the real part of the quasinormal frequencies at the best WKB order. We found that $|Im(\omega)|$ increases with the free parameter $c_{0}$ meaning a growing of the damping factor. In contrast, the real part of the quasinormal frequencies increases with $c_{0}$ implying an increasing in the oscillation of the signal. An interesting point that should be addressed before concluding this work is that for the fundamental mode, the $Im(\omega)$ approach to zero resembles what happened for the quasinormal modes of a massive scalar field on a Reissner-Nordstr\"{o}m background. It would be interesting to perform a study of the QNM thorough the WKB with Padé aproximants. However, this analysis is out of the scope of this work and we leave this and other aspects to future developments.

\subsection*{Acknowledgments}
The authors would like to acknowledge R. Konoplya for sharing his Mathematica notebook with higher order WKB corrections.
%

%
%

\bibliography{references.bib}
\bibliographystyle{unsrt}

\end{document}